\documentclass[preprint,12pt]{elsarticle}

\usepackage{amssymb}
\usepackage{amsmath}
\usepackage{booktabs}
\usepackage{multirow}
\usepackage{graphicx}
\usepackage{url}
\usepackage{subcaption}
\usepackage[colorlinks, linkcolor=black, citecolor=black, urlcolor=black]{hyperref}

\journal{Nuclear Physics B}

\begin{document}

\begin{frontmatter}



\title{Superconducting Properties of the Titanium-Based Oxides Compounds: A Review}


\author[university]{Junqi He\fnref{equal}\corref{corresponding}}
\ead{hejunqi@cjlu.edu.cn}
\author[university]{Yi Zhou\fnref{equal}\corref{corresponding}}
\ead{zhouyi43967@cjlu.edu.cn}
\affiliation[university]{organization={Department of Physics, China Jiliang University},
            city={Hangzhou},
            postcode={310018}, 
            state={Zhejiang},
            country={China}}
\fntext[equal]{The two authors contribute equally to this work.}
\cortext[corresponding]{Co-corresponding author.}
\begin{abstract}
In recent years, the superconductivity of novel layered materials, titanium-based pnictide oxides, was discovered. Due to the properties of possessing both cuprate and iron-based superconductors, these compounds have attracted the interest of researchers. Titanium pnictide oxides were reported to have CDW or SDW anomalies, theoretical calculations indicate that this DW behavior originates from the Ti$_2$O layer. These compounds which have Ti$_2$O layers provide a basis for studying the relationship between superconductivity and DW behavior. Superconductivity and DW behavior are two different electronic behaviors that typically compete with each other, but sometimes coexist. The relationship between them has always been a focus of condensed matter physics research. Through in-depth research on titanium-based superconductors, it may help us explain the unconventional superconducting transition phenomena present in iron-based superconductors. In this review, we introduce the latest research on titanium pnictide oxides and the electronic properties of these novel superconductors.
\end{abstract}

\begin{graphicalabstract}
\includegraphics[scale=0.3]{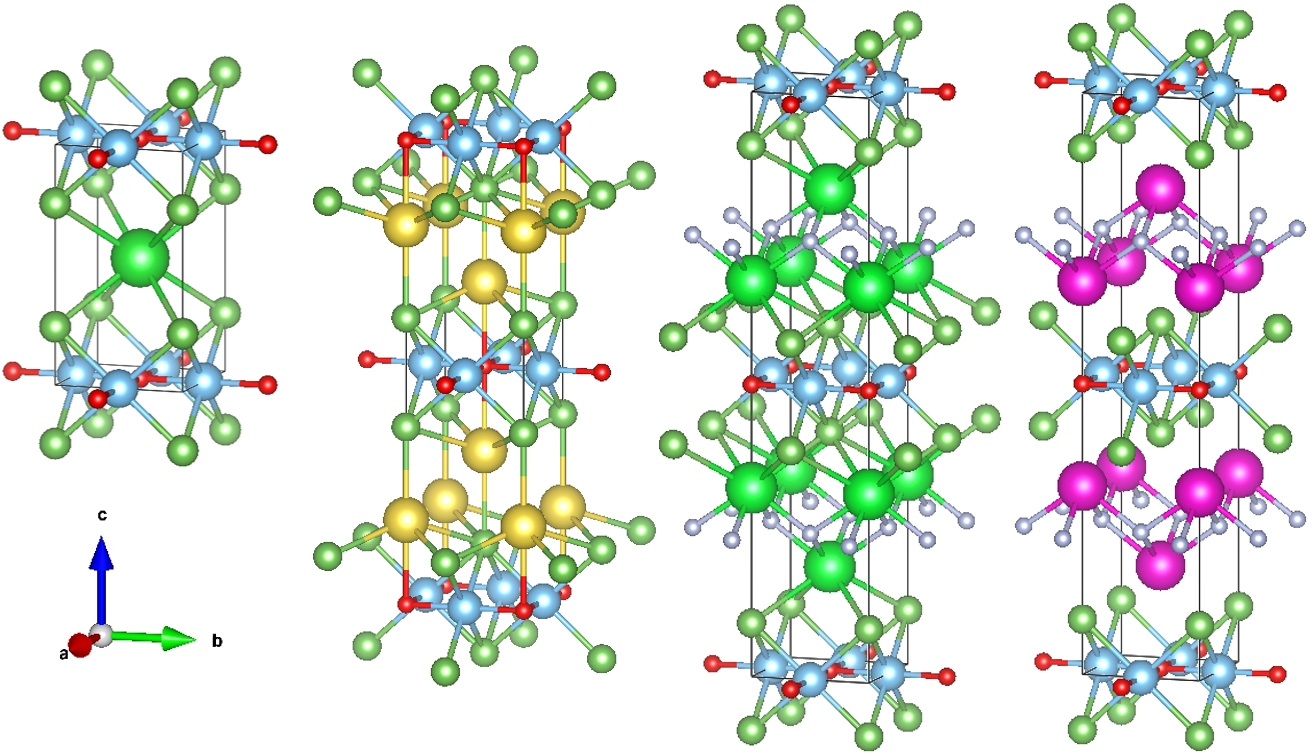}
\end{graphicalabstract}

\begin{highlights}
\item a new family of superconducting materials
\item titanium pnictide oxides possesses both the structure of cuprate superconductors and the physical properties of iron-based superconductors
\item the coexistence of conventional and unconventional superconductors provides a platform for understanding the mechanism of unconventional superconducting transitions
\end{highlights}

\begin{keyword}
superconductivity \sep titanium pnictide oxide \sep charge-density-wave \sep spin-density-wave \sep 


\end{keyword}

\end{frontmatter}



\section{Introduction}
Cuprate and iron-based superconductors are two significant branches of high-temperature superconductor research, showcasing both similarities and notable differences in their structure, superconducting mechanisms, and physical properties. Cuprate superconductors are characterized by CuO$_2$ planes, where superconductivity arises from strong electron-electron correlations in a doped Mott insulating parent phase. In contrast, iron-based superconductors are built around Fe$_2$Pn$_2$ (Pn = As)\cite{kamihara2008iron} layers, with a metallic parent phase where the suppression of spin-density-wave (SDW) order and associated residual spin fluctuations play a central role in the emergence of superconductivity. 

There is a certain connection between titanium-based superconductors and the two mentioned above, mainly reflected in their layered structure, density-wave (DW) behavior, and similarities in unconventional superconductivity mechanisms, while also having significant differences. Structurally, the titanium pnictide oxides compounds form two-dimensional Ti$_2$O layers (anti structure to the CuO$_2$ planes), capped by pnictogen ions (similar to Fe$_2$Pn$_2$ layers)\cite{lorenz2014superconductivity}. The position of Cu ions is occupied by O ions, and the position of O ions is occupied by Ti ions. The Pn (Pn = As, Sb) ions directly above and below the center of the Ti$_2$O square lattice, so {Ti}$^{3+}$ and {O}$^{2-}$ ions form a square lattice adopting the anti-CuO$_2$-type structure. Regarding the superconductivity of titanium-based pnictide oxides, they show a SDW or charge-density-wave (CDW) phase which coexists with superconductivity in some members of the family. The physical properties of those compounds are similar to iron pnictides, and different from cuprate. The parent compounds of pnictide oxides are metals with specific nesting properties of the Fermi surface which leads to the DW instability\cite{lorenz2014superconductivity}. The theoretical calculation results show that these DW behaviors are derived from the Ti$_2$O layer. The outermost electron of {Ti}$^{3+}$ is d$^{1}$ and the outermost electron of Cu$^{+2}$ is d$^{9}$. The different electronic configurations result in different electron correlations and Fermi surface properties, but their interlayer interactions are weak, so they all exhibit quasi two-dimensional electron transport properties.

In 1990, the first titanium pnictide oxide compounds Na$_2$Ti$_2$Pn$_2$O (Pn = As, Sb) was discovered, and its abnormal resistivity and magnetization are reminiscent of CDW/SDW\cite{adam1990darstellung}. From iron-based superconductors, superconductivity is generally closely related to SDW. SDW is considered to be assisted by partial nesting of quasi-two-dimensional electron and hole Fermi surface pockets\cite{scalapino2012common}, the strong spin fluctuations that still exist after suppression of the SDW are widely considered to play a central role in mediating unconventional superconductivity\cite{davies2016coupled}. In layered cuprates, CDW and superconductivity are found to be competitive\cite{miao2019formation}. The enhancement of superconductivity at the onset of CDW order in a metal, so charge fluctuations near the onset of CDW order may play an important role in the superconducting pairing mechanism\cite{wang2015enhancement}. The configuration in Na$_2$Ti$_2$Pn$_2$O and cuprate are electron-hole-symmetric. These compounds with Ti$_2$O layers lay the foundation for studying the relationship between superconductivity and DW behavior. Because of the DW states, the titanium pnictide oxides are believed to possess exotic superconductivity\cite{yajima2017titanium}. Until 2012, a research team discovered a novel layered superconductor BaTi$_2$Sb$_2$O ($T_c$ = 1.2 K) and density wave coexist near 50 K\cite{yajima2012superconductivity}. At the same time, the coexistence of superconductivity ($T_c$ = 21 K) and 125 K density wave was reported in Ba$_2$Ti$_2$Fe$_2$As$_4$O containing FeAs layer and Ti$_2$O layer. But the superconductivity of this compound still comes from the FeAs layer, not from the Ti$_2$O layer\cite{sun2012ba2ti2fe2as4o}. The following year (2013), superconductor BaTi$_2$Bi$_2$O was discovered by the same research team\cite{yajima2012synthesis}. The superconductor of titanium pnictide oxides compounds BaTi$_2$Pn$_2$O (Pn = Sb, Bi) consist of superconducting Ti$_2$Pn$_2$O layers and Ba blocking layers. In addition to the above compounds, many compounds have been synthesized such as Na$_2$Ti$_2$Pn$_2$O (Pn = As, Sb)\cite{liu2009physical}, (SrF)$_2$Ti$_2$Pn$_2$O (Pn = As, Sb, Bi)\cite{yajima2012synthesis,liu2010structure}, (SmO)$_2$Ti$_2$Sb$_2$O\cite{liu2010structure}, (EuF)$_2$Ti$_2$Pn$_2$O (Pn = Sb, Bi)\cite{zhai2022structure} and BaTi$_2$Pn$_2$O (Pn = As, Sb, Bi)\cite{yajima2013two}, but only BaTi$_2$2Pn$_2$O (Pn = Sb, Bi) shows superconductivity.

In this review, the author summarizes some research results and introduces the physical properties of titanium pnictide oxides compounds and the superconducting research.

\section{Na$_2$Ti$_2$Pn$_2$O (Pn = As, Sb)}
Layered cuprates and pnictides have high superconducting transition temperatures and some abnormal physical properties, such as antiferromagnetism or CDW/SDW\cite{scalapino2012common, sadovskii2008high,machida1980theory,morosan2006superconductivity}. These abnormal physical properties are closely related to the unconventional superconductivity. So through research, Na$_2$Ti$_2$Pn$_2$O was proposed and synthesized\cite{adam1990darstellung}. But Na$_2$Ti$_2$Pn$_2$O have no superconductivity.

\begin{figure}[h]
    \centering
    \includegraphics[scale=0.25]{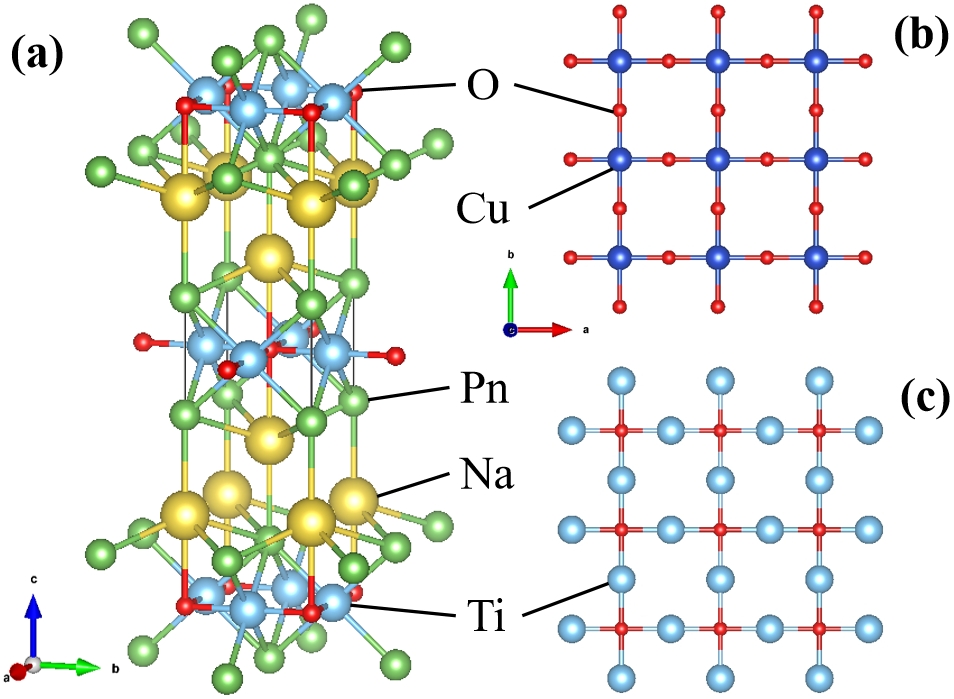}
    \caption{(a) Crystal structure of Na$_2$Ti$_2$Pn$_2$O (Pn = As ,Sb). (b) CuO2 square lattice in La$_2$CuO$_4$. (c) Ti$_2$O square lattice in Na$_2$Ti$_2$Pn$_2$O.}
    \label{1}
\end{figure}

Na$_2$Ti$_2$Pn$_2$O is an anti-K$_2$NiF$_4$-type structure\cite{liu2009physical}, which is similar to the structure of La$_2$CuO$_4$\cite{longo1973structure}. This structure is similar to the anti-CuO$_2$-type structure observed in cuprate. As shown in Figure \ref{1}a, crystallizes of Na$_2$Ti$_2$Pn$_2$O in $I4/mmm$ symmetry and edge-shared layers, in which double layers of Na$^{+}$ are interspersed\cite{yajima2017titanium,liu2009physical}. Ti$^{3+}$ cation is located between two O$^{2-}$ ions forming a square plane layer of Ti$_2$O, and two Pn$^{3-}$ ions are located above and below the center of the square unit of Ti$_2$O. These ions are octahedral coordination to form the Ti$_2$Pn$_2$O layer. [Ti$_2$Pn$_2$O]$^{2-}$ unit consisting of a Ti-O layer with an anticonfiguration to the CuO$_2$ layer. In Figure \ref{1}b and c, Ti and O replace O and Cu in the square lattice respectively. 

The space group of Na$_2$Ti$_2$Pn$_2$O are $I4/mmm$. The lattice constants are $a=4.079$ {\AA} and $c=15.26$ {\AA} for Na$_2$Ti$_2$As$_2$O and $a=4.153$ {\AA} and $c=16.57$ {\AA} for Na$_2$Ti$_2$Sb$_2$O\cite{liu2009physical}. From the perspective of structure, because the interlayer distance of Na$_2$Ti$_2$Sb$_2$O is larger than that of Na$_2$Ti$_2$As$_2$O, Na$_2$Ti$_2$Sb$_2$O has a more two-dimensional nature (stronger anisotropy) than Na$_2$Ti$_2$As$_2$O\cite{miao2019formation}. The CDW/SDW instability is suppressed as the dimensionality of the system increases. Therefore, the CDW/SDW amplitude in Na$_2$Ti$_2$Sb$_2$O is more obvious\cite{liu2009physical, ozawa2001possible}.

\begin{figure}[h]
    \centering
    \includegraphics[scale=0.25]{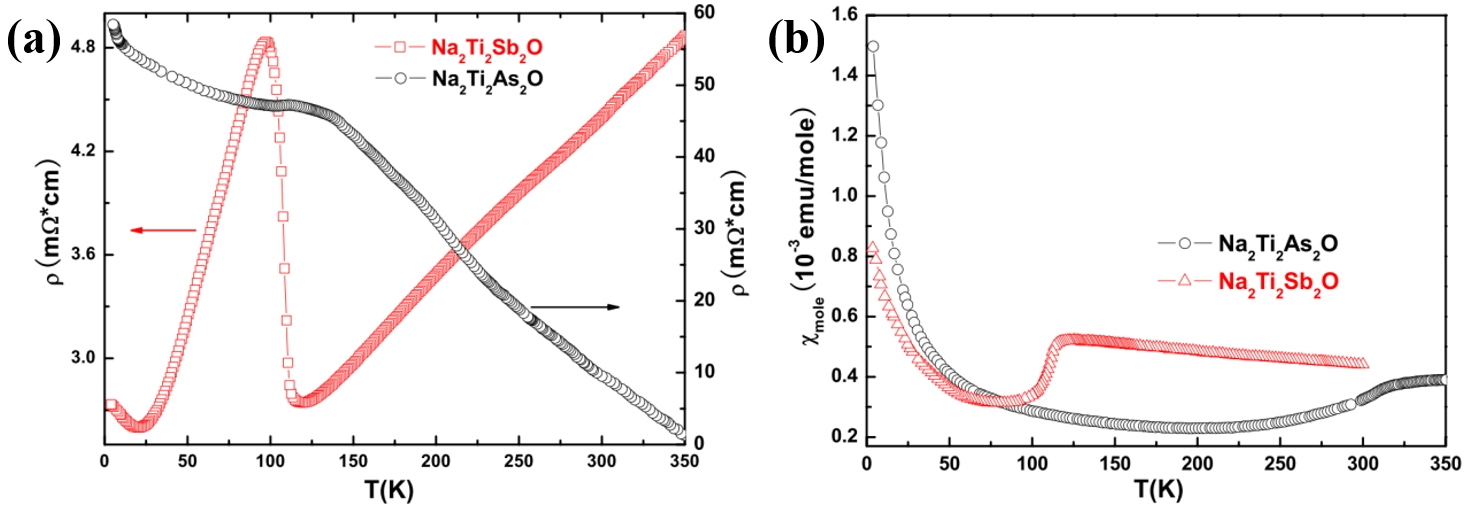}
    \caption{(a) Temperature dependence of resistivity for the Na$_2$Ti$_2$Pn$_2$O (Pn = As ,Sb). (b) Temperature dependence of magnetic susceptibility measured under 5 T for Na$_2$Ti$_2$Pn$_2$O in the temperature range of 4-300 K\cite{liu2009physical}.}
    \label{2}
\end{figure}

The relationship between the resistivity and temperature is shown in Figure \ref{2}a, the resistivity of Na$_2$Ti$_2$Sb$_2$O exhibits an anomalous transition at about 115 K, the resistivity increases sharply at about 115 K. The resistivity of Na$_2$Ti$_2$As$_2$O has no obvious anomalous transition within the measured temperature range, but a weak kink is observed at about 135 K. In Figure \ref{2}b, the magnetic susceptibility of Na$_2$Ti$_2$Sb$_2$O exhibits an anomalous transition at approximately 115 K, which is consistent with the anomaly in resistivity. The magnetic susceptibility of Na$_2$Ti$_2$Sb$_2$O decreases sharply at about 115 K. The magnetic susceptibility of Na$_2$Ti$_2$As$_2$O also decreases at about 320K, but the resistivity is not abnormal. The anomalous transition in magnetic susceptibility and resistivity of Na$_2$Ti$_2$Pn$_2$O are very similar to that observed in high-$T_c$ iron-based oxypnictide compounds, such as LnOFeAs (1111) and AEFe$_2$As$_2$ (122)\cite{ivanovskii2009new}. In these materials, powder neutron diffraction studies have revealed a structural phase transition from a tetragonal to an orthorhombic lattice, along with the formation of SDW, which explains the observed anomalies in magnetic susceptibility and resistivity\cite{scalapino2012common,zhao2009spin}.

\begin{figure}[h]
    \centering
    \includegraphics[scale=0.25]{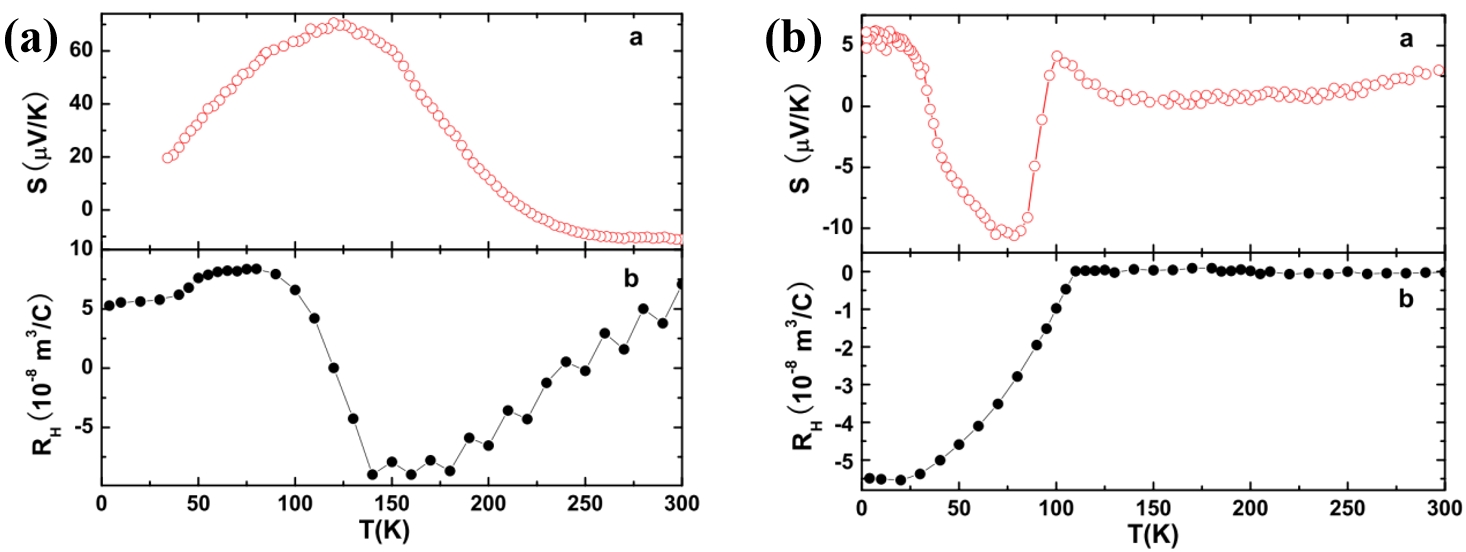}
    \caption{(a) Temperature dependence of thermoelectric power (top) and Hall coefficient (lower) for Na$_2$Ti$_2$As$_2$O. (b) Temperature dependence of thermoelectric power (top) and Hall coefficient (lower) for Na$_2$Ti$_2$Sb$_2$O\cite{liu2009physical}.}
    \label{3}
\end{figure}

In Figure \ref{3}, it shows the temperature dependence of thermoelectric power and Hall coefficient for Na$_2$Ti$_2$Pn$_2$O. The Hall coefficient of Na$_2$Ti$_2$Bi$_2$O is small and negative and shows a weak temperature dependence above 115 K. This mainly indicates that the dominant carrier is electrons. The Hall coefficient increases significantly below 115 k and the magnetic susceptibility and resistivity also have anomaly at same temperature. This indicates that the decrease in carrier concentration below 115 K may be due to the opening of the CDW/SDW gap\cite{liu2009physical}. This behavior also exists in SmFeAsO$_{1-x}$F$_x$\cite{PhysRevLett.101.087001}, mainly due to strong antiferromagnetism (or SDW) fluctuations exist. The thermoelectric power of Na$_2$Ti$_2$Bi$_2$O decreases sharply at about 110 K, which is also consistent with the abnormal transition observed in resistivity, magnetic susceptibility and Hall coefficient. This further indicates that there is an abnormality in the carrier concentration at the occurrence temperature of CDW/SDW ordering. It should be mentioned that above 110 K, the Hall coefficient and thermoelectric power show contrasting sign, indicating a multiple-bands behavior in Na$_2$Ti$_2$Bi$_2$O\cite{liu2009physical, kuchinskii2009multiple,tanaka2001phase}. In addition, the $^{23}$Na nuclear magnetic resonance measurements of Na$_2$Ti$_2$Pn$_2$O shows that the phase transition does not change nuclear magnetic resonance spectra structure. It indicates that there is no ordered magnetic moment, which excludes the possibility of SDW instability\cite{fan2013charge}. And other experiments also show the possibility of CDW phase transition\cite{fan2013charge,PhysRevB.94.104515}.

The Hall coefficient of Na$_2$Ti$_2$As$_2$O decreases monotonically and changes the sign from positive to negative with decreasing temperature to about 135 K. Then the Hall coefficient increases at about 135 K. The Hall coefficient is positive below 100 K, indicating a weak temperature dependence.\cite{liu2009physical} The thermoelectric power exhibits a broad peak at 125 K and increases below 125 K and decreases above 125 K. At about 200 K, the thermoelectric power changes the sign from positive to negative. The Theoretical calculations indicate that complex phase transitions arise from CDW transitions driven by the Fermi surface nestings\cite{PhysRevB.96.155142}. Na$_2$Ti$_2$As$_2$O shows different physical properties from Na$_2$Ti$_2$Bi$_2$O. Na$_2$Ti$_2$As$_2$O has little abnormal transition at the occurrence temperature of CDW/SDW ordering.

\begin{figure}[h]
    \centering
    \includegraphics[scale=0.23]{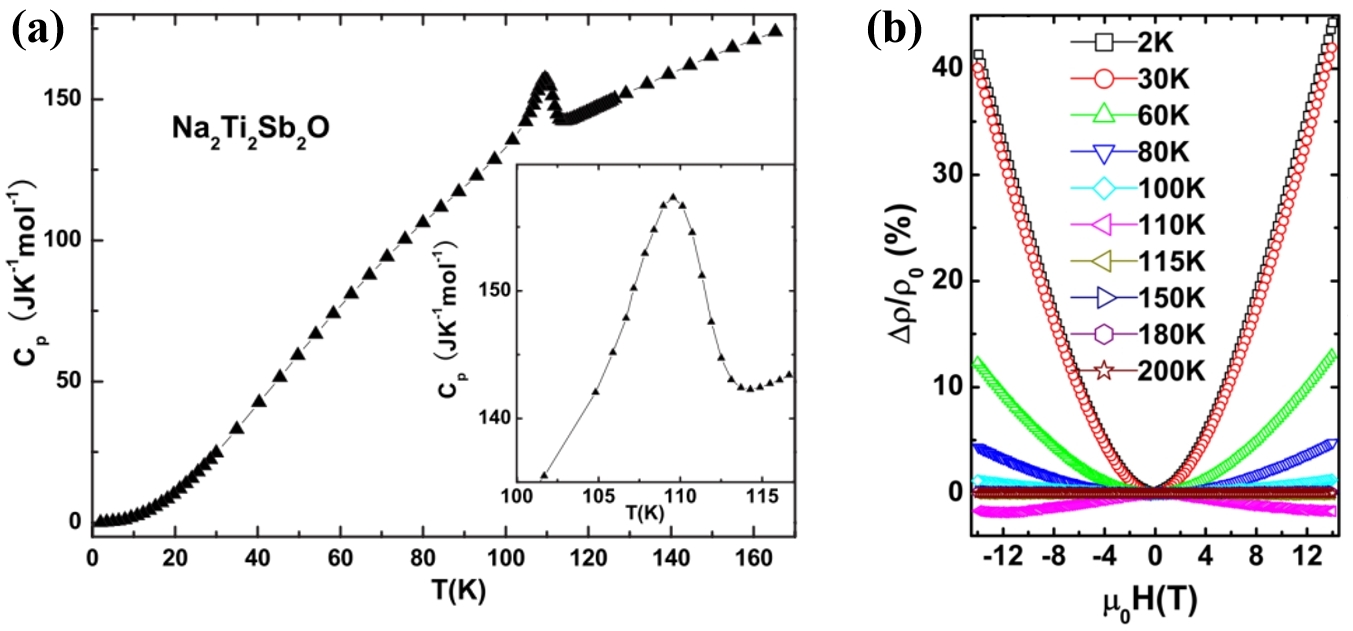}
    \caption{(a) Temperature dependence of specific heat for Na$_2$Ti$_2$Sb$_2$O. Inset shows the $C_p/T$ vs $T$ curve around $T_s$. (b) Magnetic field dependence of magnetoresistance for Na$_2$Ti$_2$Sb$_2$O at different temperatures.\cite{yajima2017titanium,liu2009physical}.}
    \label{4}
\end{figure}

Figure \ref{4}a shows the temperature dependence of the specific heat for Na$_2$Ti$_2$Sb$_2$O. It is obvious that there is an abnormal peak at about 110 K. This clearly indicates that a phase transition occurred at 110 K approximately, which may have been caused by the CDW/SDW phase transition\cite{liu2009physical}. In Figure \ref{4}b, it can be observed that a larger magnetoresistance was found in Na$_2$Ti$_2$Sb$_2$O. It is considered to be the suppression of SDW instability by the magnetic field in Na$_2$Ti$_2$Sb$_2$O\cite{yajima2017titanium}. In the low-temperature domain, the specific heat of the material is accurately represented by the expression $C_p=\gamma T+\beta T^3$. The Debye temperature $\Theta_D$ is determined by the formula $\beta=(12/5NR\pi^4)/({\Theta_D}^3)$, where $N$ represents the number of atoms per formula unit. The Sommerfeld coefficients is estimated to be $\gamma$ = 28 mJ·K$^{-2}$·mol$^{-1}$, $\beta$ = 1.49 mJ·K$^{-4}$·mol$^{-1}$, and the Debye temperature $\Theta_D$ is calculated to be 209 K for Na$_2$Ti$_2$Sb$_2$O\cite{liu2009physical}.

Some researchers synthesized Na$_2$Ti$_2$Sb$_2$O$_{1-x}$F$_x$ ($x=0.00-0.50$)\cite{doan2012syntheses}. After F doping, the amount of drop of SDW transition is reduced, but the superconducting transition was still not observed. In addition, there has been no significant progress in the doping of Na$_2$Ti$_2$Pn$_2$O. Due to the instability of Na cations in Na$_2$Ti$_2$Pn$_2$O, it is very difficult to doping carriers into Na$_2$Ti$_2$Pn$_2$O to suppress SDW and introduce possible superconductivity\cite{Wang_2010}. Although Na$_2$Ti$_2$Pn$_2$O is not superconducting, some of its physical properties can be used to explain why some phenomena occur. The synthesis of Na$_2$Ti$_2$Bi$_2$O still needs to be studied, and there is currently no successful synthesis. 

\section{(SrF)$_2$Ti$_2$Pn$_2$O (Pn = As, Sb, Bi) and (SmO)$_2$Ti$_2$Sb$_2$O}

\begin{figure}[h]
    \centering
    \includegraphics[scale=0.23]{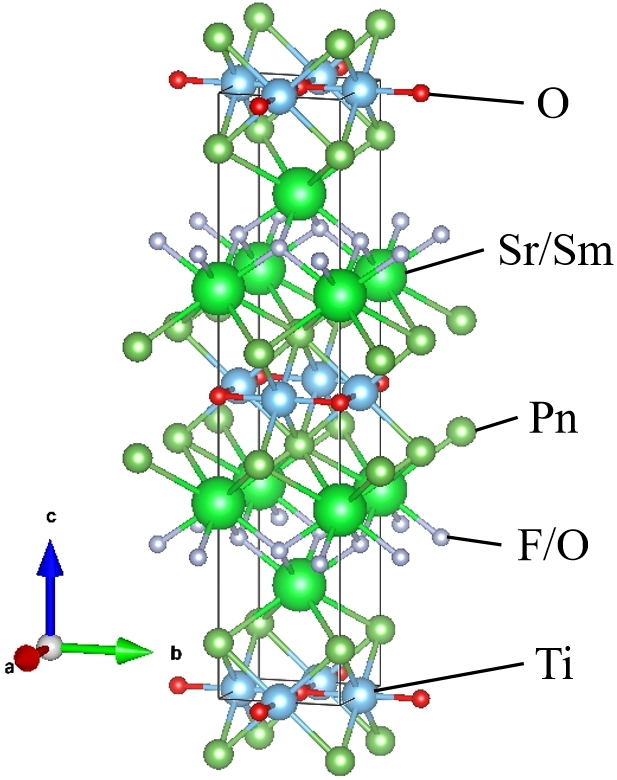}
    \caption{Crystal structures of (SrF)$_2$Ti$_2$Pn$_2$O (Pn = As, Sb, Bi) and (SmO)$_2$Ti$_2$Sb$_2$O (Sr and Sm are in the same position, F and O are in the same position).}
    \label{5}
\end{figure}

(SrF)$_2$Ti$_2$Pn$_2$O (Pn = As, Sb) and (SmO)$_2$Ti$_2$Sb$_2$O\cite{liu2010structure} were proposed in 2009 and (SrF)$_2$Ti$_2$Bi$_2$O\cite{yajima2012synthesis} were proposed in 2013. (SrF)$_2$Ti$_2$Pn$_2$O (Pn = As, Sb) and (SmO)$_2$Ti$_2$Sb$_2$O are predicted by the structure and physical properties of Na$_2$Ti$_2$Pn$_2$O\cite{liu2010structure}. Na$_2$Ti$_2$Pn$_2$O is an anti-K$_2$NiF$_4$-type structure. Therefore, as long as the double layers of Na$^{+}$ are replaced by a fluorite-type [Sr$_2$F$_2$]$^{2+}$ or [Sm$_2$O$_2$]$^{2+}$ layers, as shown in Figure \ref{5}. (SrF)$_2$Ti$_2$Bi$_2$O was proposed in the same way, which is the first Pn = Bi compound proposed in the ATi$_2$Pn$_2$O family (BaTi$_2$Bi$_2$O was also found at the same time). But its physical properties are different from (SrF)$_2$Ti$_2$Pn$_2$O (Pn = As, Sb).

The space groups of (SrF)$_2$Ti$_2$Pn$_2$O and (SmO)$_2$Ti$_2$Sb$_2$O are $I4/mmm$. Lattice constants for the tetragonal unit cell are determined to be $a=4.049$ {\AA} and $c=19.42$ {\AA} for (SrF)$_2$Ti$_2$As$_2$O, $a=4.110$ {\AA} and $c=20.89$ {\AA} for (SrF)$_2$Ti$_2$Sb$_2$O, $a=4.118$ {\AA} and $c=21.37$ {\AA} for (SrF)$_2$Ti$_2$Bi$_2$O, $a=4.032$ {\AA} and $c=20.15$ {\AA} for (SmO)$_2$Ti$_2$Sb$_2$O. (SmO)$_2$Ti$_2$Sb$_2$O are slightly less than that of (SrF)$_2$Ti$_2$Sb$_2$O because the thickness of the SmO layer is less than that of the SrF layer\cite{yajima2012synthesis,liu2010structure}.

\begin{figure}[h]
    \centering
    \includegraphics[scale=0.25]{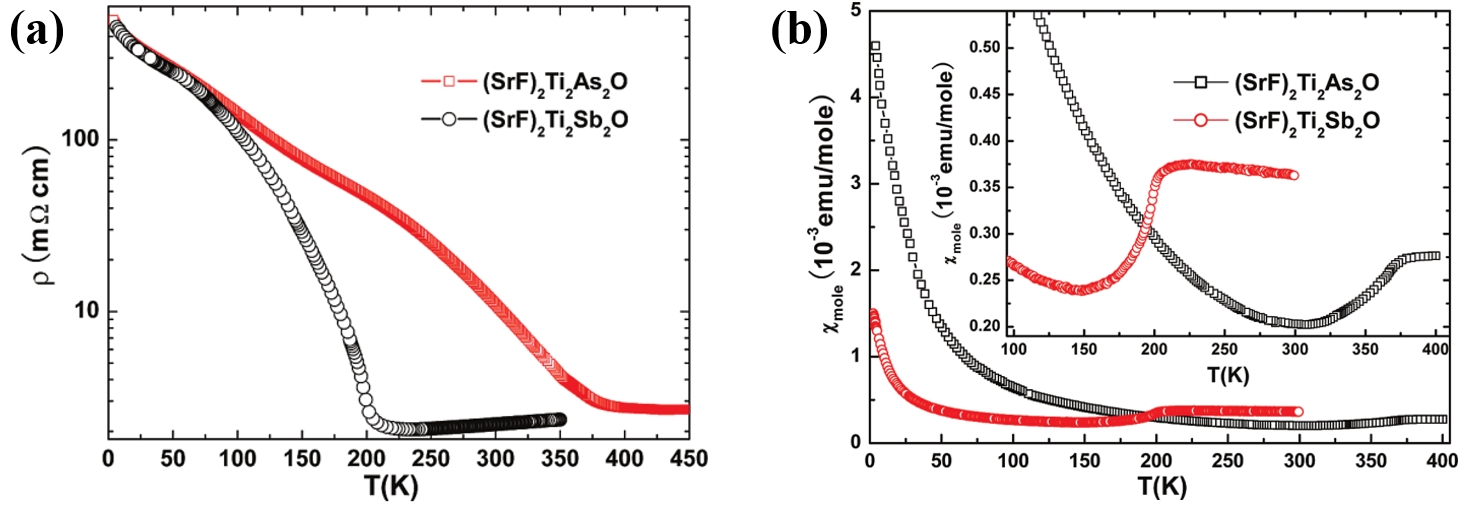}
    \caption{(a) Temperature dependence of resistivity for (SrF)$_2$Ti$_2$Pn$_2$O (Pn = As, Sb). (b) Temperature dependence of magnetic susceptibility measured under 5 T for (SrF)$_2$Ti$_2$Pn$_2$O (Pn = As, Sb). Inset shows the $\chi$ vs $T$ curve around $T_s$\cite{liu2010structure}.}
    \label{6}
\end{figure}

Figure \ref{6}a shows temperature dependence of resistivity for (SrF)$_2$Ti$_2$Pn$_2$O (Pn = As, Sb). The resistivity of (SrF)$_2$Ti$_2$Pn$_2$O (Pn = As, Sb) exhibits a metal-to-insulator transition at about 200 K and 380 K respectively, and the resistivity increases sharply at a certain temperature, which should be attributed to the opening of a gap. Figure \ref{6}b shows the magnetic susceptibility measured at 5 T, the susceptibility of (SrF)$_2$Ti$_2$Sb$_2$O decreased sharply at about 200 K, and a Curie paramagnetic tail at low temperature. The magnetic susceptibility of (SrF)$_2$Ti$_2$As$_2$O decreased sharply at about 380 K. It can be observed that both resistivity and magnetic susceptibility exhibit abnormal transitions at the same temperature.

\begin{figure}[h]
    \centering
    \includegraphics[scale=0.25]{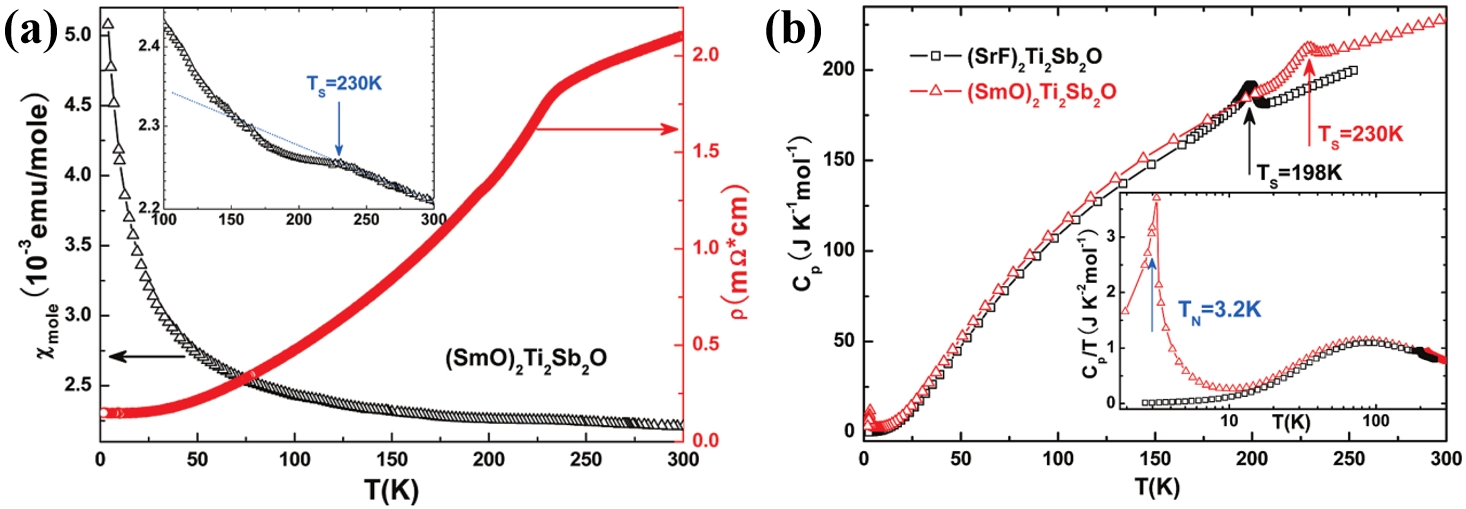}
    \caption{(a) Temperature dependence of resistivity (red) and magnetic susceptibility (black) measured under 5 T for (SmO)$_2$Ti$_2$Sb$_2$O. Inset shows the $\chi$ vs $T$ curve around $T_s$. (b) Temperature dependence of specific heat for (SrF)$_2$Ti$_2$Sb$_2$O and (SmO)$_2$Ti$_2$Sb$_2$O. Inset shows the sharp peak comes from the antiferromagnetic ordering of Sm$^{3+}$ ions in $C_p/T$ vs $\ln T$ curve\cite{liu2010structure}.}
    \label{7}
\end{figure}

The anomalous transition in resistivity and magnetic susceptibility also obtains in (SmO)$_2$Ti$_2$Sb$_2$O. In Figure \ref{7}a, the resistivity of (SmO)$_2$Ti$_2$Sb$_2$O exhibits an anomaly at 230 K, but this is different from the metal-to-insulator transition of (SrF)$_2$Ti$_2$Pn$_2$O (Pn = As, Sb). And in inset of Figure \ref{7}a, the susceptibility of (SmO)$_2$Ti$_2$Sb$_2$O also exhibits a minor anomaly at about 230 K. These behaviors are similar to those observed in Na$_2$Ti$_2$Pn$_2$O. Temperature dependence of specific heat for (SrF)$_2$Ti$_2$Sb$_2$O and (SmO)$_2$Ti$_2$Sb$_2$O can be seen from Figure \ref{7}b. There is an obvious abnormal peak at 198 K for (SrF)$_2$Ti$_2$Sb$_2$O and 230 K for (SmO)$_2$Ti$_2$Sb$_2$O. Lattice distortion and instability of CDW/SDW may lead to significant peaks. It can also be seen from the inset of Figure \ref{7}b that (SmO)$_2$Ti$_2$Sb$_2$O has another abnormal peak at 3.2 K. This specific heat behavior also obtains in the high-$T_C$ cuprate. It is proved that the antiferromagnetic ordering of Sm$^{3+}$ ions in Sm$_2$CuO$_4$ at 5.9 K\cite{hayashi2015modulation,takada2004influences}. In addition, there are many compounds that have this phenomenon, and these compounds have the same structure of a fluorite-type [Sm$_2$O$_2$]$^{2+}$ layers, so the peak at 3.2 K in (SmO)$_2$Ti$_2$Sb$_2$O obviously manifests the antiferromagnetic ordering of Sm$^{3+}$ ions. By fitting the low-temperature region, the Sommerfeld coefficients of (SrF)$_2$Ti$_2$Sb$_2$O and (SmO)$_2$Ti$_2$Sb$_2$O can be obtained as $\gamma$ = 7.73 mJ·K$^{-2}$·mol$^{-1}$, $\beta$ = 1.13 mJ·K$^{-4}$·mol$^{-1}$ and $\gamma$ = 170.15 mJ·K$^{-2}$·mol$^{-1}$, $\beta$ = 0.72 mJ·K$^{-4}$·mol$^{-1}$. The calculated Debye temperature $\Theta_D$ for (SrF)$_2$Ti$_2$Sb$_2$O and (SmO)$_2$Ti$_2$Sb$_2$O are 249k and 289k, respectively.

\begin{figure}
    \centering
    \includegraphics[scale=0.25]{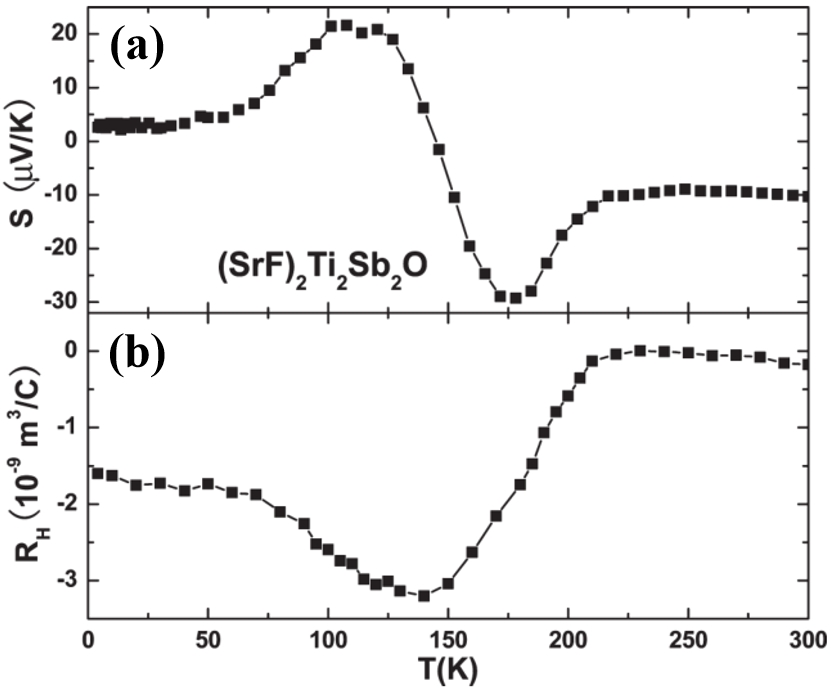}
    \caption{(a) Temperature dependence of thermoelectric power (a) and Hall coefficient (b) for (SrF)$_2$Ti$_2$Sb$_2$O.curve\cite{liu2010structure}.}
    \label{8}
\end{figure}

In addition to these physical properties mentioned above, (SrF)$_2$Ti$_2$Sb$_2$O has some abnormal phenomena. In Figure \ref{8}a, it can be observed that the thermoelectric power and Hall coefficient of (SrF)$_2$Ti$_2$Sb$_2$O exhibit significant anomalous transitions around 200 K, both showing strong temperature dependence, indicating that the conductivity mechanism of the material may involve multiple energy bands and competitive behavior of charge carriers. The positive and negative changes in Hall coefficient indicate the competitive behavior between electrons and holes, which may be caused by the opening of the CDW/SDW gap leading to a decrease in charge carriers\cite{liu2010structure}. These behaviors are very similar to iron-based oxypnictide compounds\cite{wang2009effects}. In fact, by measuring the lattice constants of (SrF)$_2$Ti$_2$Sb$_2$O at different temperatures, it can be found that the sample exhibits significant lattice distortion at 200 K\cite{liu2010structure}. Due to lattice distortion, anomalies occurred in resistivity, susceptibility, Hall coefficient and specific heat.

\begin{figure}
    \centering
    \includegraphics[scale=0.3]{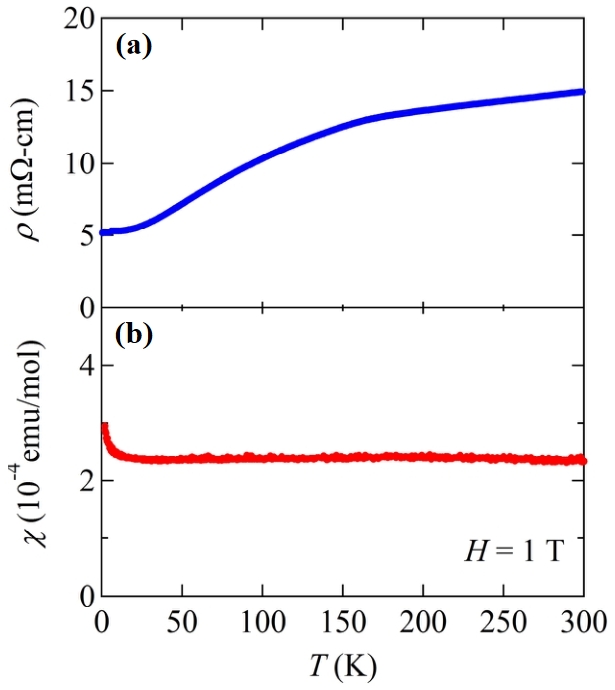}
    \caption{(a) Temperature dependence of resistivity for (SrF)$_2$Ti$_2$Bi$_2$O. (b) Temperature dependence of magnetic susceptibility for (SrF)$_2$Ti$_2$Bi$_2$O at a magnetic field of 1 T\cite{yajima2017titanium}.}
    \label{9}
\end{figure}

The resistivity and magnetic susceptibility curves of (SrF)$_2$Ti$_2$Bi$_2$O are shown in Figure \ref{9}. (SrF)$_2$Ti$_2$Bi$_2$O is different from (SrF)$_2$Ti$_2$Sb$_2$O, it is no obvious abnormality associated with a CDW/SDW transition. The resistivity has no abnormality and the susceptibility increases very slightly with decreasing temperature\cite{yajima2012synthesis}. The metallic temperature dependence was observed in (SrF)$_2$Ti$_2$Bi$_2$O over the whole temperature range measured. These results show that replacing the As and Sb site with Bi can completely inhibit the CDW/SDW instability. But (SrF)$_2$Ti$_2$Bi$_2$O is not superconducting and there is no diamagnetic signal even below 0.5 K.

\section{BaTi$_2$Pn$_2$O (Pn = As, Sb, Bi)}
BaTi$_2$Pn$_2$O (Pn = As, Sb, Bi) is the only compounds in the family of titanium pnictide oxides that has found superconductivity, and superconductivity is only observed in BaTi$_2$Sb$_2$O and BaTi$_2$Bi$_2$O. These three compounds were synthesized in 2009 (BaTi$_2$As$_2$O\cite{Wang_2010}) and 2012 (BaTi$_2$Sb$_2$O\cite{yajima2012superconductivity}, BaTi$_2$Bi$_2$O\cite{yajima2012synthesis}) respectively. Although BaTi$_2$Pn$_2$O (Pn = Sb, Bi) is superconducting, it is not a high-$T_c$ superconductor like cuprates.

\begin{figure}[h]
    \centering
    \includegraphics[scale=0.3]{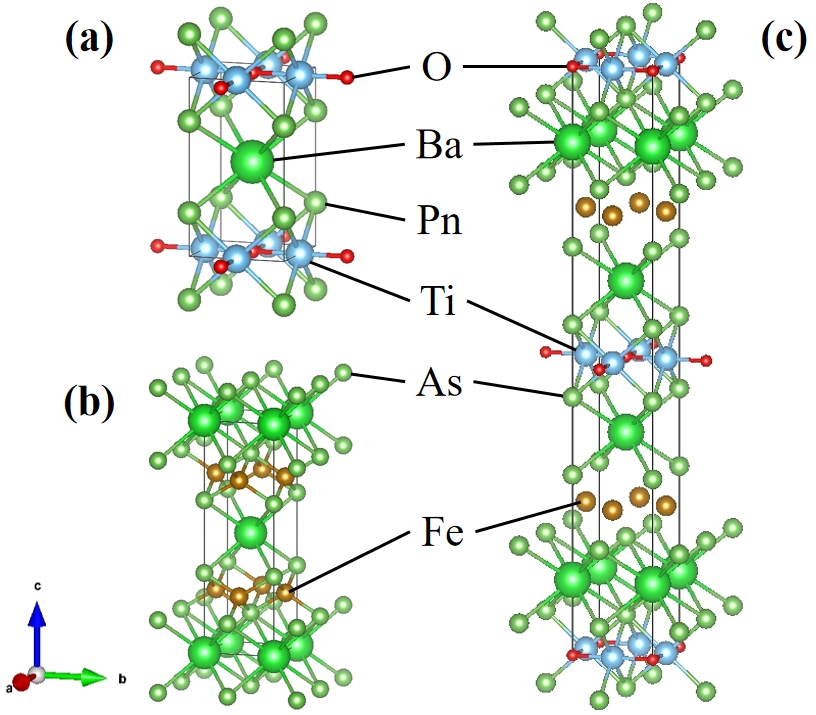}
    \caption{Crystal structures of (a) BaTi$_2$Pn$_2$O (Pn = As, Sb, Bi), (b) BaFe$_2$As$_2$ and (c) Ba$_2$Ti$_2$Fe$_2$As$_4$O.}
    \label{10}
\end{figure}

The structure of BaTi$_2$Pn$_2$O is very similar to that of Na$_2$Ti$_2$Pn$_2$O. As shown in Figure \ref{10}a, two [Ti$_2$Pn$_2$O]$^{2-}$ layers with single-layer Ba$^{2+}$ inside. The double layer Na$^{+}$ was replaced by the single layer Ba$^{2+}$, [Ti$_2$Pn$_2$O]$^{2-}$ layers and Ba$^{2+}$ layers are stacked alternately along the $c$-axis\cite{Wang_2010}. It is very obvious that the length of the $c$-axis of BaTi$_2$Pn$_2$O is smaller than that of Na$_2$Ti$_2$Pn$_2$O. Because the intermediate layer (four layers Na$^{+}$ and single layer [Ti$_2$Pn$_2$O]$^{2-}$) of Na$_2$Ti$_2$Pn$_2$O is not present in center of the BaTi$_2$Pn$_2$O unit cell and BaTi$_2$Pn$_2$O only have one single layer Ba$^{2+}$ inside, the length of the $c$-axis of BaTi$_2$Pn$_2$O is only about half of that of Na$_2$Ti$_2$Pn$_2$O. Ba$_2$Ti$_2$Fe$_2$As$_4$O can be regarded as a compound derived from BaTi$_2$Pn$_2$O. The crystal structure of Ba$_2$Ti$_2$Fe$_2$As$_4$O (Figure \ref{10}c) can be regarded as a combination of BaFe$_2$As$_2$\cite{jiang2013crystal} (Figure \ref{10}b)  and BaTi$_2$As$_2$O. Ba$_2$Ti$_2$Fe$_2$As$_4$O contains a Ti$_2$O layer, but the main reason for its superconductivity is the Fe$_2$As$_2$ layer\cite{sun2012ba2ti2fe2as4o}. The Fe$_2$X$_2$ layer is considered to be the crucial structural unit of iron-based superconductivity\cite{dong2008competing,ren2009research}. So Ba$_2$Ti$_2$Fe$_2$As$_4$O is essentially an iron-based superconductor, not titanium-based superconductor.

The space groups of BaTi$_2$Pn$_2$O are $P4/mmm$. This is different from other titanium pnictide oxides. The lattice constants of As, Sb, and Bi compounds are $a=4.047$ {\AA} and $c=7.275$ {\AA}, $a=4.110$ {\AA} and $c=8.086$ {\AA}, $a=4.123$ {\AA} and $c=8.345$ {\AA}, respectively\cite{yajima2012superconductivity,yajima2012synthesis,Wang_2010}. It can be observed that the $c$-axis of BaTi$_2$Sb$_2$O and BaTi$_2$Bi$_2$O is longer than that of BaTi$_2$As$_2$O, mainly because the atomic radius of Pn gradually increases from As to Bi, thus elongating the $c$-axis.

\begin{figure}[h]
    \centering
    \includegraphics[scale=0.38]{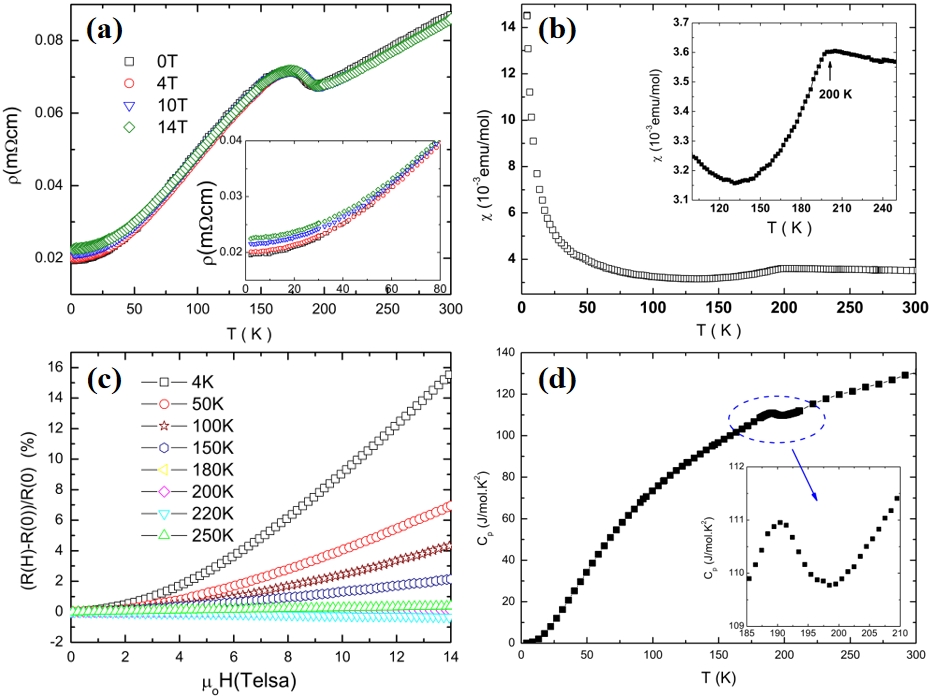}
    \caption{(a) Temperature dependence of resistivity measured at different magnetic fields for BaTi$_2$As$_2$O (the inset shows the low temperature part of the $R–T$ curve at different magnetic fields). (b) Temperature dependence of susceptibility measured under 1 T for BaTi$_2$As$_2$O (the inset shows the data zoom in near the phase transition). (c) Isothermal magnetoresistance at different temperatures for BaTi$_2$As$_2$O. (d) Temperature dependence of specific heat for BaTi$_2$As$_2$O (the inset shows the anomaly of specific heat around the phase transition)\cite{Wang_2010}.}
    \label{11}
\end{figure}

BaTi$_2$As$_2$O is the only one in this family that does not exhibit superconductivity. From Figure \ref{11}a, it can be seen that the resistivity of BaTi$_2$As$_2$O generally increases with temperature, exhibiting typical metallic behavior, which is completely opposite to that of Na$_2$Ti$_2$As$_2$O. As shown in Figure \ref{11}b, the magnetic susceptibility exhibits weak temperature dependence above 200 K. There is a significant abnormal transition in resistivity and susceptibility around 200 K. The reasons for these anomalies may be structural distortion and CDW/SDW. It can be observed that below 200 K, BaTi$_2$As$_2$O exhibits significant magnetic resistance (Figure \ref{11}c), and as the temperature decreases, the magnetic resistance increases. This indicates the SDW ordering of BaTi$_2$As$_2$O at 200 K. At the same time, there was an abnormal jump in specific heat at 200 K (Figure \ref{11}d), indicating that due to the CDW/SDW transition, BaTi$_2$As$_2$O had a bulk phase transition at approximately 200 K.

To induce superconductivity in BaTi$_2$As$_2$O, doping was performed. Due to the relatively small radius of Li$^+$, it can be doped into the gap, thus increasing the concentration of Li$^+$ can suppress the instability of CDW/SDW and induce superconducting transition. In BaLi$_x$Ti$_{2-x}$As$_2$O, the maximum value is reached when $x=0.2$, and the instability of CDW/SDW is indeed suppressed, but no superconducting transition occurs\cite{Wang_2010}. Similarly, Cr doping can suppress the instability of CDW/SDW (BaCr$_x$Ti$_{2-x}$As$_2$O), but no superconducting transition was observed above 2 K\cite{ji2014synthesis}.

\begin{figure}[h]
    \centering
    \includegraphics[scale=0.35]{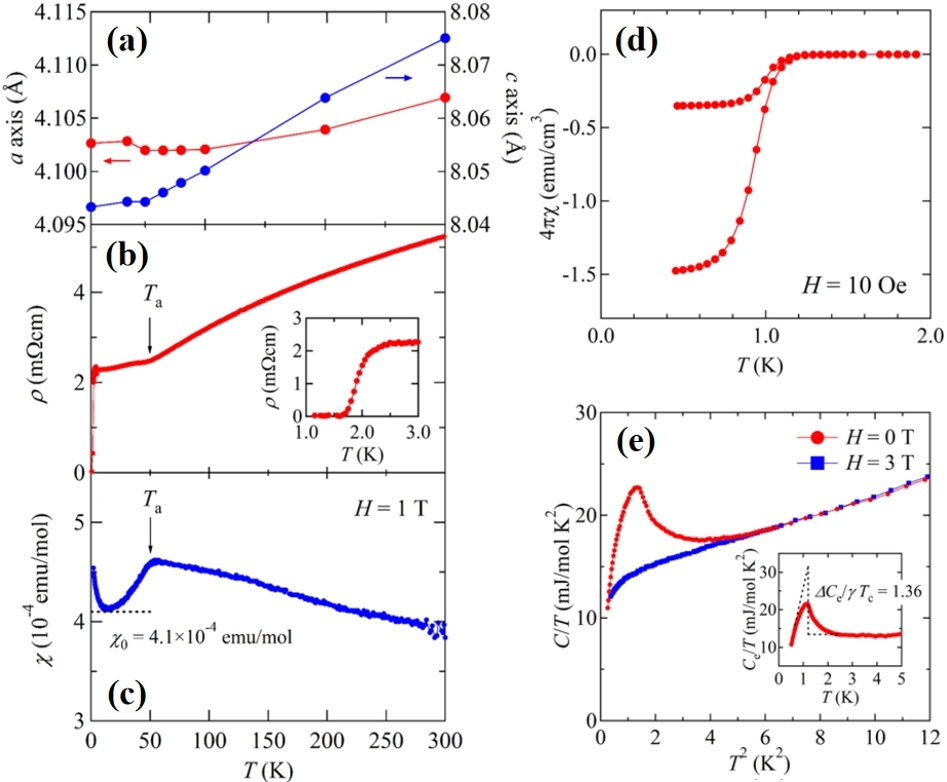}
    \caption{(a) Temperature dependence of lattice constants for BaTi$_2$Sb$_2$O (the red line represents the $a$-axis and the blue line represents the $c$-axis). (b) Temperature dependence of resistivity for BaTi$_2$Sb$_2$O (the inset shows the data zoom in near the phase transition). (c) Temperature dependence of magnetic susceptibility measured under 1 T for BaTi$_2$Sb$_2$O. (d) Low-temperature magnetic susceptibility at 10 Oe for BaTi$_2$Sb$_2$O. (e) Temperature dependence of specific heat for BaTi$_2$Sb$_2$O (the inset shows a fitting plot of electronic specific heat $C_e$)\cite{yajima2017titanium,yajima2012superconductivity}.}
    \label{12}
\end{figure}

In BaTi$_2$Sb$_2$O, it can be observed that there is an abnormal transition at approximately 50 K, as shown in Figure \ref{12}a, b, c. At $T_a$, significant abnormalities were observed in lattice constants, resistivity, and magnetic susceptibility, and the transition temperature was much lower than that of BaTi$_2$As$_2$O. These abnormal transformations are also largely caused by the CDW/SDW transition. At 2 K, the resistance drops sharply to 0, and at 0.8 K, a diamagnetic signal appears (as shown in Figure \ref{12}d). From Figure \ref{12}e, it can also be seen that there is a clear peak in specific heat at approximately 0.8 K. These phenomena demonstrate the bulk superconductivity in BaTi$_2$Sb$_2$O. By estimating the Sommerfeld coefficient, it can be found that the values of BaTi$_2$As$_2$O and BaTi$_2$Sb$_2$O are approximate (BaTi$_2$As$_2$O: $\gamma$ = 15.3 mJ·K$^{-2}$·mol$^{-1}$, $\beta$ = 0.79 mJ·K$^{-4}$·mol$^{-1}$, $\Theta_D$ = 223 K\cite{Wang_2010}. BaTi$_2$Sb$_2$O: $\gamma$ = 13.5 mJ·K$^{-2}$·mol$^{-1}$, $\beta$ = 0.857 mJ·K$^{-4}$·mol$^{-1}$, $\Theta_D$ = 239 K\cite{yajima2012superconductivity}). According to entropy conservation, 1.36 are estimated from the specific heat jump, which is almost consistent with the BCS weak-coupling limit of 1.43\cite{yajima2012superconductivity}.

\begin{figure}[h]
    \centering
    \includegraphics[scale=0.28]{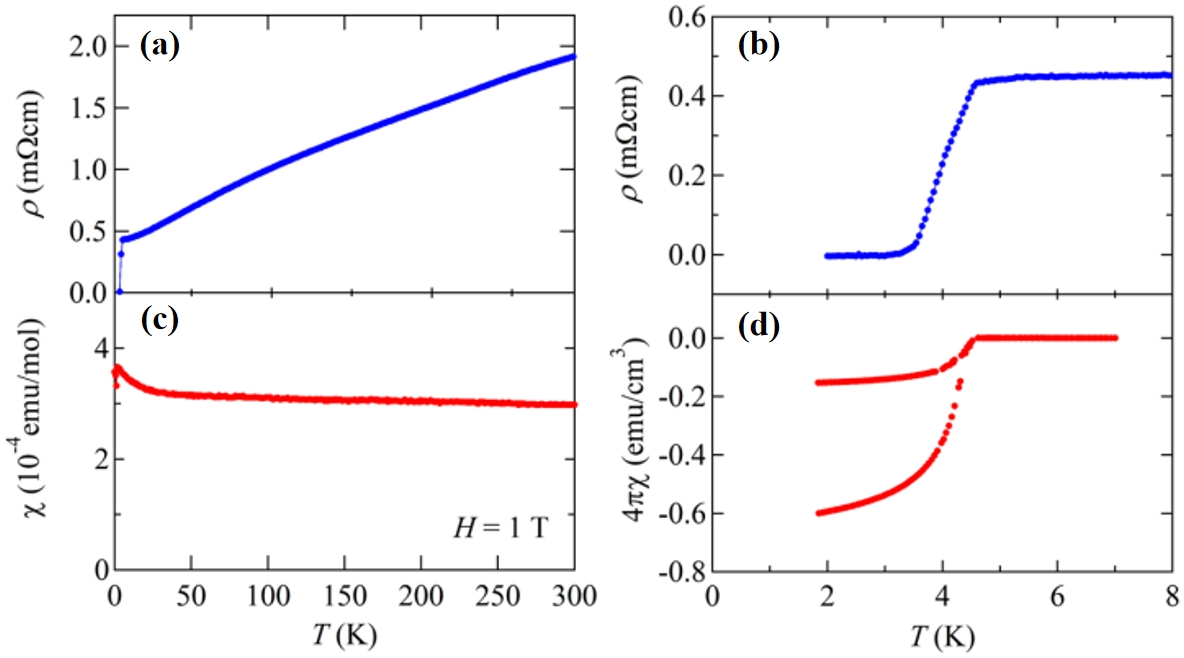}
    \caption{(a) Temperature dependence of resistivity for BaTi$_2$Bi$_2$O. (b) Resistivity in low-temperature regions. (c)Temperature dependence of magnetic susceptibility measured under 1 T for BaTi$_2$Bi$_2$O. (d) Magnetic susceptibility in low-temperature regions\cite{yajima2017titanium,yajima2012synthesis}.}
    \label{13}
\end{figure}

BaTi$_2$Bi$_2$O is completely different from BaTi$_2$As$_2$O and BaTi$_2$Sb$_2$O, and there is no significant abnormal transformation in resistivity and magnetic susceptibility. As shown in the Figure \ref{13}a, BaTi$_2$Bi$_2$O exhibits metal temperature dependence throughout the entire temperature measurement range. It exhibits paramagnetism in terms of magnetic susceptibility, similar to As and Sb. No obvious DW behavior was found within the measurement range. These results indicate that replacing As and Sb sites with Bi can completely suppress the instability of CDW/SDW. At 4.6 K, BaTi$_2$Bi$_2$O exhibits superconducting transition, with a sharp decrease in resistance to 0 and the appearance of diamagnetic signals (Figure \ref{13}b). Due to the suppression of the instability of CDW/SDW, the superconducting transition temperature of BaTi$_2$Bi$_2$O is significantly higher than that of BaTi$_2$Sb$_2$O.

\begin{figure}[h]
    \centering
    \includegraphics[scale=0.33]{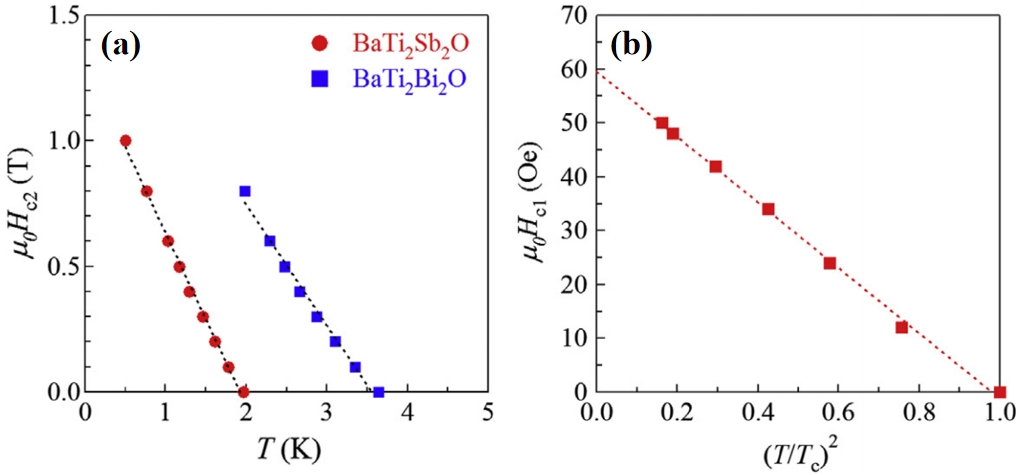}
    \caption{(a) The upper critical field $H_{c2}(0)$ for BaTi$_2$Sb$_2$O and BaTi$_2$Bi$_2$O in low-temperature regions. (b) The lower critical field $H_{c1}(0)$ for BaTi$_2$Bi$_2$O as a function of $(T/Tc)^2$ in low-temperature regions\cite{yajima2014superconducting}.}
    \label{14}
\end{figure}

Although BaTi$_2$Sb$_2$O and BaTi$_2$Bi$_2$O are both superconductors, there are significant differences in their superconducting properties. As shown in Figure \ref{14}a, the upper critical field of the two superconductors is linearly related to temperature. By fitting the upper critical field, it can be found that at 0 K, the $H_{c2}(0)$ of BaTi$_2$Sb$_2$O is 1.32 T, and the $H_{c2}(0)$ of BaTi$_2$Bi$_2$O is 1.7 T. According to $H_{c2}(0)$, the coherence lengths $\xi_0$ of the two superconductors are estimated to be 158 \AA and 139 \AA, respectively. The $H_{c2}(0)$ of BaTi$_2$Sb$_2$O is low and the coherence length $\xi_0$ is long, indicating that the material is at the BCS weak-coupling limit, with relatively weak electron-phonon coupling, making it a conventional BCS-type superconductor\cite{monthoux1992weak,PhysRevB.87.060510}. The relatively high $H_{c2}(0)$ of BaTi$_2$Bi$_2$O indicates that superconductivity can remain stable even under high magnetic fields. By fitting, the lower critical field $H_{c1}(0)$ of BaTi$_2$Bi$_2$O was found to be 60 Oe (Figure \ref{14}b). Combined with other superconducting parameters, it was estimated that the density of states of BaTi$_2$Bi$_2$O is smaller than that of BaTi$_2$Sb$_2$O. This indicates that the reason for the increase in superconducting transition temperature cannot be explained under the BCS weak-coupling limit, and BaTi$_2$Bi$_2$O is not a conventional superconductor\cite{yajima2017titanium,yajima2014superconducting}.

\begin{figure}[h]
    \centering
    \includegraphics[scale=0.33]{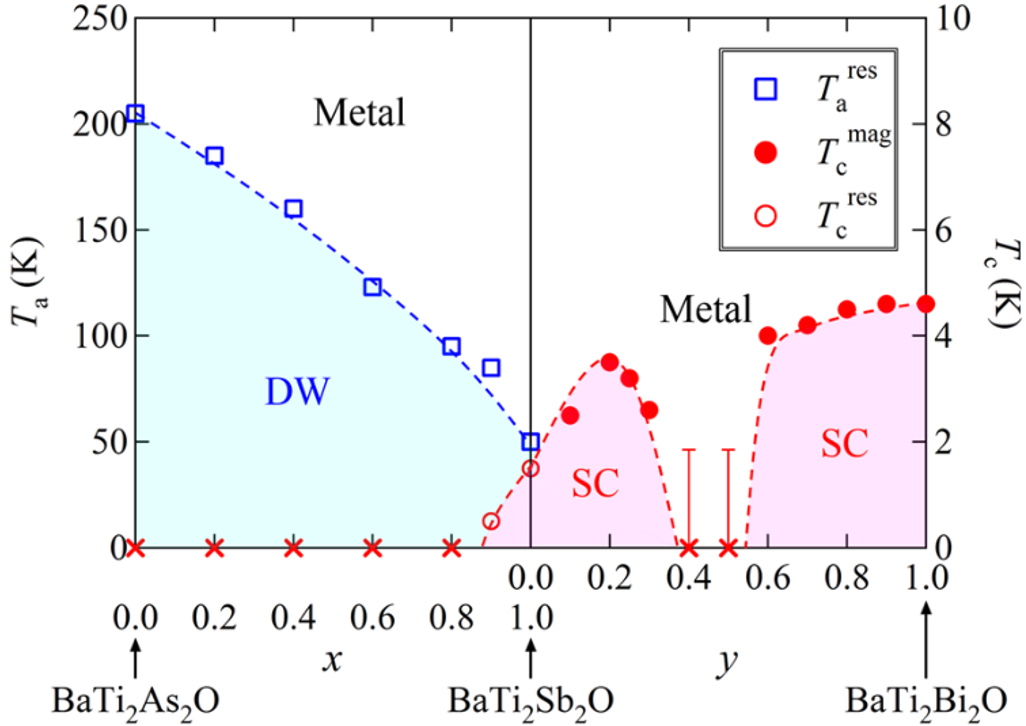}
    \caption{(a) The upper critical field $H_{c2}(0)$ for BaTi$_2$Sb$_2$O and BaTi$_2$Bi$_2$O in low-temperature regions. (b) The lower critical field $H_{c1}(0)$ for BaTi$_2$Bi$_2$O as a function of $(T/Tc)^2$ in low-temperature regions\cite{yajima2014superconducting}.}
    \label{15}
\end{figure}

Study the reasons for the increase in $T_c$ by doping Pn. By synthesizing BaTi$_2$(As$_{1-x}$Sb$_x$)$_2$O and BaTi$_2$(Sb$_{1-x}$Bi$_x$)$_2$O, it was discovered that that there is a two-dome structure in the $T_c$ as shown in Figure \ref{15}a\cite{yajima2013two,PhysRevB.98.220507}. These two domes represent two superconducting phases respectively. It can be found that the first dome range is in $0.9 \leq x \leq 1$ and $0 \leq y \leq 0.3$, then the maximum $T_c$ = 3.5 K occurs when $y=0.2$. The second dome range is in $0.6 \leq y \leq 1$, then the maximum $T_c$ = 4.6 K occurs when $y=1$. No superconductivity was detected in the area between the two domes above 1.85 K\cite{yajima2013two,PhysRevB.98.220507}. It can be found that the shape of the two domes is different. The first dome presents a peak, while the $T_c$ of the second dome rises with the increase of $y$. In the first dome, superconductivity is mainly due to the suppression of CDW/SDW instability\cite{yajima2013two}. For the second dome, because of the phenomenon of no superconductivity in the middle region and the sudden rise of $T_c$, it means that it has different superconductivity mechanism from the first dome. This phenomenon also exists in the iron-based superconductor LaFeAsO$_{1-x}$H$_x$\cite{PhysRevB.89.094510}, mainly due to the electronic doping. This type of compound with two superconducting phases typically exhibits strong multi-band properties on the Fermi surface. So BaTi$_2$(Sb$_{1-x}$Bi$_x$)$_2$O has multi-band properties\cite{ji2014synthesis}.

\begin{figure}
    \centering
    \includegraphics[scale=0.3]{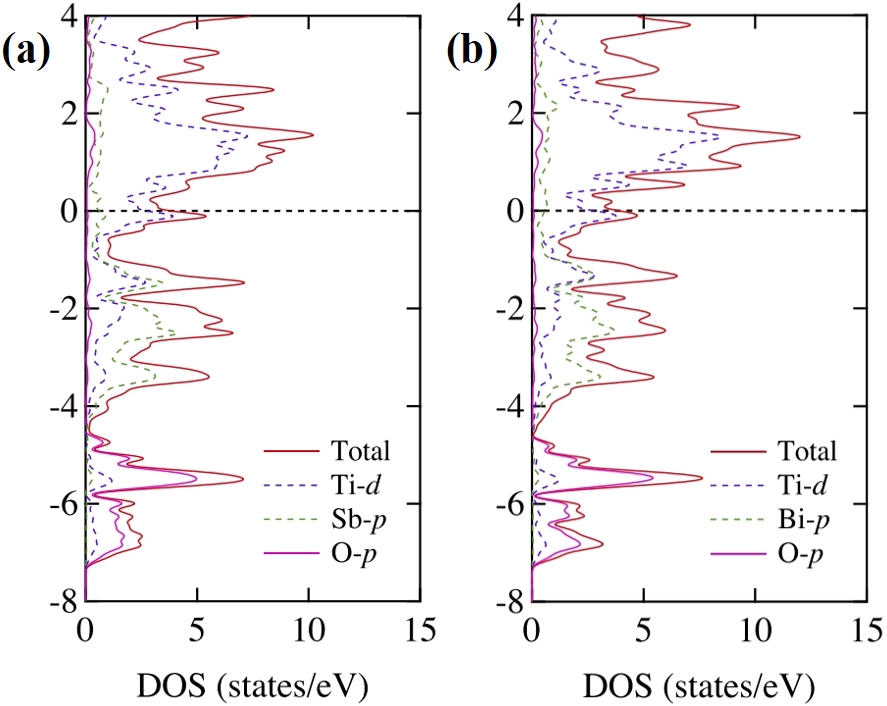}
    \caption{Densities of states for (a) BaTi$_2$Sb$_2$O and (b) BaTi$_2$Bi$_2$O\cite{nakano2016phonon}.}
    \label{16}
\end{figure}

\begin{figure}
    \centering
    \includegraphics[scale=0.23]{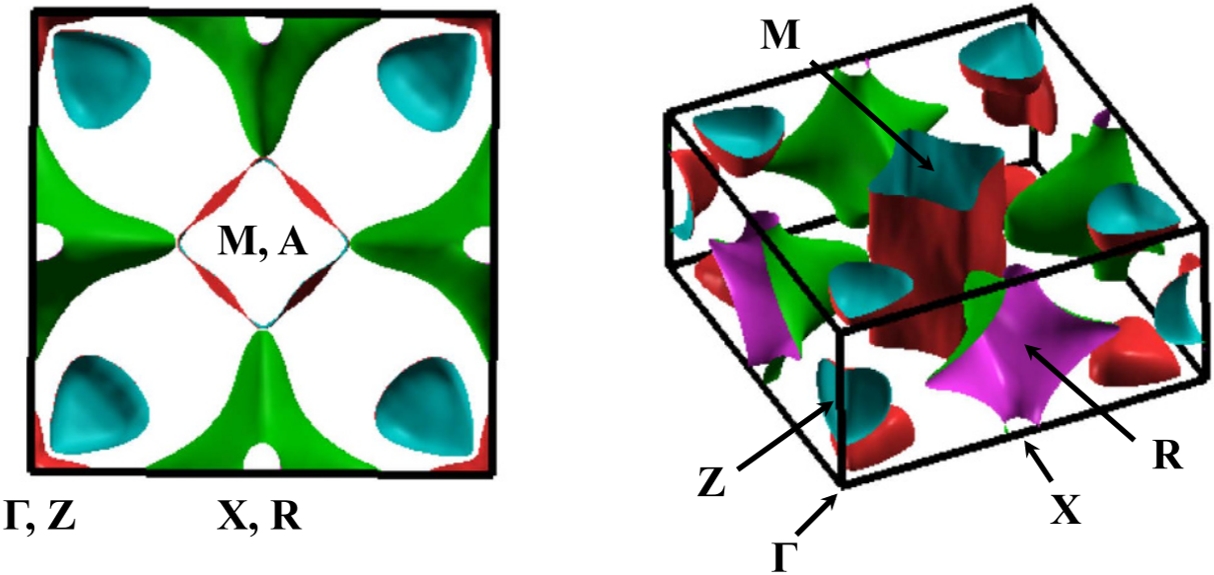}
    \caption{Fermi surfaces of BaTi$_2$Bi$_2$O\cite{nakano2016phonon}.}
    \label{17}
\end{figure}

Through first-principles calculations, it can be found that BaTi$_2$Sb$_2$O and BaTi$_2$Bi$_2$O do exhibit multi-band properties, with Ti-3$d$ orbitals ($d_{xy}$, $d_{x^2-y^2}$, $d_{z^2}$) making the greatest contribution to the Fermi level ($E_F$)\cite{yajima2017titanium,suetin2013electronic,PhysRevB.89.155108,singh2012electronic,frandsen2014intra,PhysRevB.93.245122}. Ti-3$d$ are hybridized with Sb-5$p$ and Bi-6$p$ bands closer to the center of the Brillouin zone. From the Figure \ref{16}, it can be seen that they provide the majority of the density of states (DOS) near the Fermi energy and define the features of the Fermi surface\cite{suetin2013electronic,nakano2016phonon}. The calculated Fermi surface of BaTi$_2$Bi$_2$O is shown in Figure \ref{17}, and it can be seen that the Fermi surface is very complex. The square cylinder $M, A$ in the figure mainly surrounds the mixture of $d_{xy}$, $d_{x^2-y^2}$, $d_{z^2}$ orbitals, and this square cylinder exhibits nested property. The entire Brillouin zone also contains multiple hole pockets and electron pockets. Inhibition of CDW/SDW instability through electron or hole doping induces superconducting transition. From As to Bi doping, the Fermi level of BaTi$_2$Pn$_2$O first increases and then decreases, which is different from the change in $T_c$, so there is no correlation between Fermi level and $T_c$, and it cannot be explained by the BCS theory. This indicates that the increase in superconducting transition temperature may be due to structural distortion, and the reduced overlap between larger Pn ions and Ti-Ti atomic orbitals may suppress the instability of CDW/SDW\cite{suetin2013electronic}. 

\begin{figure}[h]
    \centering
    \includegraphics[scale=0.4]{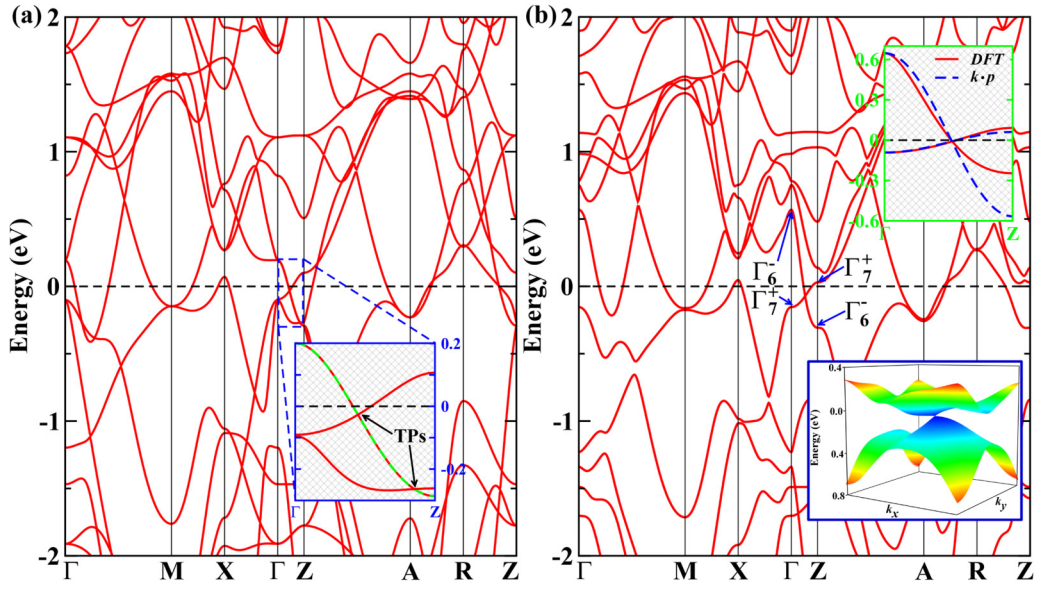}
    \caption{(a) The energy band structure of BaTi$_2$Bi$_2$O without SOC. (b) The energy band structure of BaTi$_2$Bi$_2$O with SOC (the lower inset shows the structure of a three-dimensional Dirac cone in the band)\cite{PhysRevB.99.155155}.}
    \label{18}
\end{figure}

By measuring the superconducting transition temperature of BaTi$_2$Bi$_2$O under high pressure, it was found that $T_c$ monotonically increases to 7 K before the pressure increases to 4.0 GPa, and then reaches saturation. The unit cell volume of BaTi$_2$Bi$_2$O linearly decreases under pressure, providing evidence for structural distortion leading to an increase in $T_c$. However, BaTi$_2$Bi$_2$O under pressure did not undergo a significant structural phase transition, so this phenomenon of $T_c$ increase may also be caused by electronic changes or phonon softening\cite{wang2020superconducting}. Through experiments under high pressure, it was also found that the superconducting phase of BaTi$_2$Bi$_2$O at 7.28 GPa would be explained by the $p$-wave polar model\cite{wang2020superconducting}. And because the $p$-wave superconductors is topologically nontrivial, the superconductivity of BaTi$_2$Bi$_2$O is topologically nontrivial and unconventional\cite{qin2023two}. So BaTi$_2$Bi$_2$O may be a Dirac semimetal. Based on the DFT, BaTi$_2$Bi$_2$O also conforms to the above properties. There exists a pair of three-dimensional Dirac points in the Brillouin zone as shown in Figure \ref{18}\cite{wang2020superconducting,PhysRevB.99.155155}.

\begin{figure}
    \centering
    \includegraphics[scale=0.35]{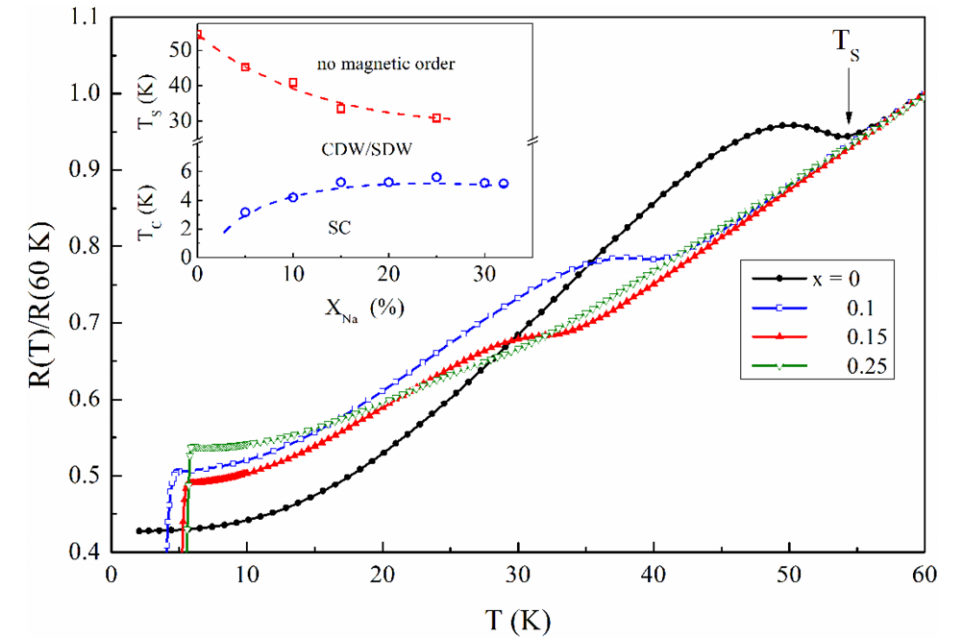}
    \caption{Normalized resistance of Ba$_{1-x}$Na$_x$Ti$_2$Sb$_2$O near the CDW/SDW transition (the inset shows the phase diagram derived from resistivity and magnetization measurements)\cite{doan2012ba1}.}
    \label{19}
\end{figure}

\begin{figure}
    \centering
    \includegraphics[scale=0.28]{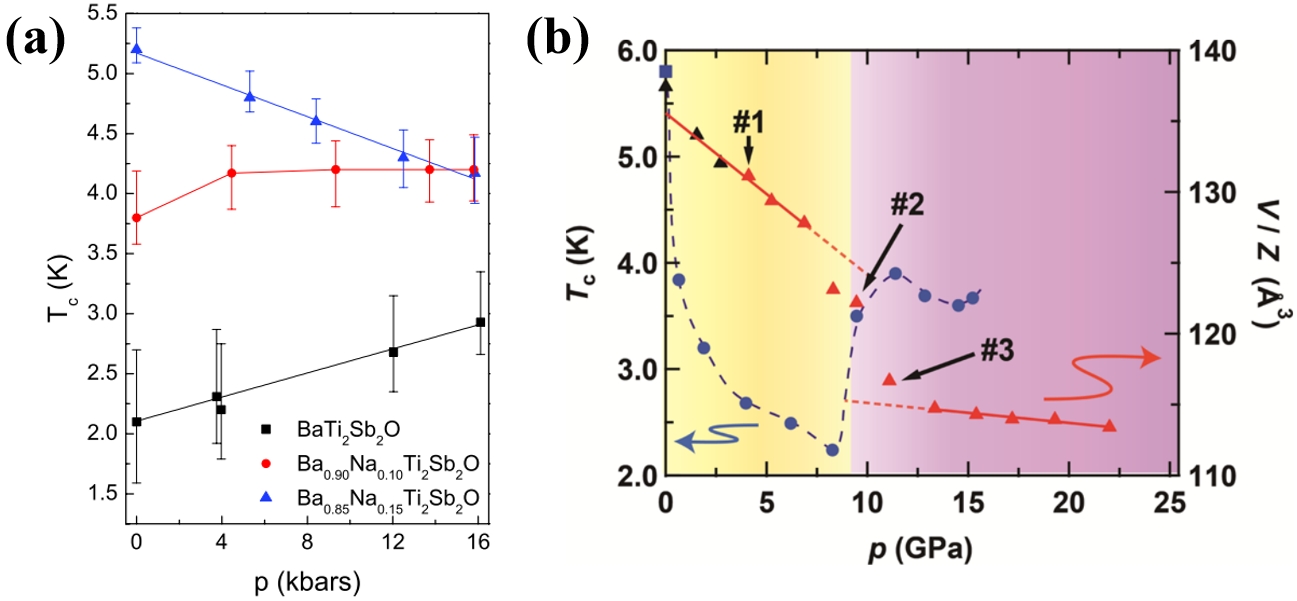}
    \caption{(a) Pressure dependence of $T_c$ for Ba$_{1-x}$Na$_x$Ti$_2$Sb$_2$O ($x=$ 0, 0.1, 0.15). (b) Pressure dependence of $T_c$ (blue line) and unit cell volume (red line) for Ba$_{0.77}$Na$_0.23$Ti$_2$Sb$_2$O\cite{gooch2013high,taguchi2021emergence}.}
    \label{20}
\end{figure}

In addition to using physical methods to increase $T_c$, chemical doping methods are also used to increase $T_c$. Among them, Ba$_{1-x}$Na$_x$Ti$_2$Sb$_2$O is the most widely studied doping material. From the Figure \ref{19}, it can be seen that doping Na can systematically suppress the instability of CDW/SDW, thereby inducing superconductivity. When the doping concentration $x$ reaches 0.15, the superconducting transition temperature can increase to 5.5 K\cite{doan2012syntheses,doan2012ba1}. By testing Ba$_{1-x}$Na$_x$Ti$_2$Sb$_2$O under pressure, it was found that as $x$ increases, the increase in $T_c$ changes from positive correlation to negative correlation with increasing pressure, as shown in Figure \ref{20}a\cite{gooch2013high}. When $x$ increased to 0.23, abnormal changes were observed in $T_c$, which monotonically decreased with increasing pressure below 8 GPa, suddenly increased rapidly at 8 GPa, and then slowly decreased again at 11 GPa (Figure \ref{20}b). This abnormal jump may be caused by the emergence of new superconducting phases\cite{taguchi2021emergence}. The superconductivity of Ba$_{1-x}$Na$_x$Ti$_2$Sb$_2$O is significantly affected by doping, pressure, and changes in crystal structure. Through calculation, it can also be found that Ba$_{1-x}$Na$_x$Ti$_2$Sb$_2$O in the Uemura scheme is on the edge of unconventional superconductivity, indicating that it may have unconventional superconductivity\cite{kamusella2014cdw}. In addition to Ba$_{1-x}$Na$_x$Ti$_2$Sb$_2$O, doping at the Ba site also includes (Ba$_{1-x}$Rb$_x$)Ti$_2$Sb$_2$O ($x=0.12$, $T_c=6.1$ K)\cite{PhysRevB.89.094505}, (Ba$_{1-x}$K$_x$)Ti$_2$Sb$_2$O ($x=0.2$, $T_c=5.4$ K)\cite{pachmayr2014superconductivity} and (Ba$_{1-x}$Cs$_x$)Ti$_2$Sb$_2$O ($x=0.25$, $T_c=4.4$ K)\cite{wang2019preparation,PhysRevB.104.184519}. And the Pn site also includes BaTi$_2$(Sb$_{1-x}$Sn$_x$)$_2$O ($x=0.3$, $T_c=2.5$ K)\cite{nakano2013t}. By doping BaTi$_2$Pn$_2$O, $T_c$ was significantly increased.

\section{(EuF)$_2$Ti$_2$Pn$_2$O (Pn = Sb, Bi)}

\begin{figure}
    \centering
    \includegraphics[scale=0.23]{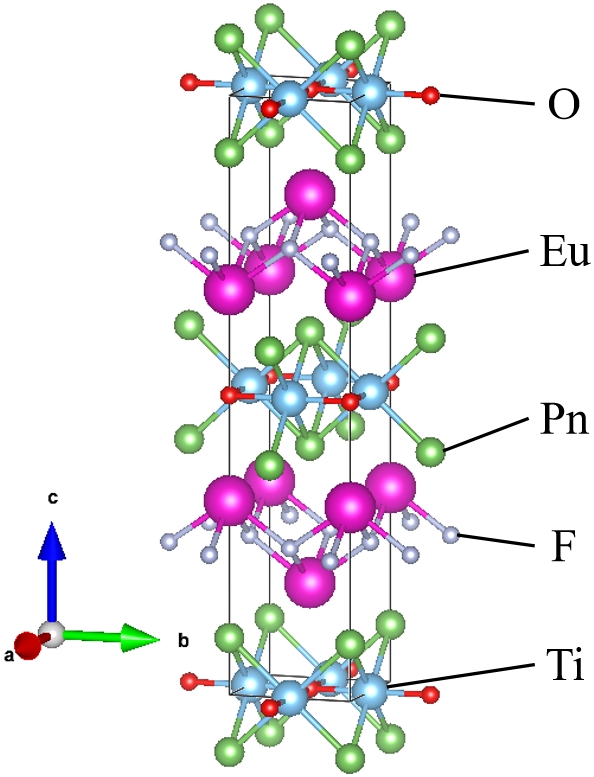}
    \caption{Crystal structures of (EuF)$_2$Ti$_2$Pn$_2$O (Pn = Sb, Bi).}
    \label{21}
\end{figure}

(EuF)$_2$Ti$_2$Pn$_2$O (Pn = Sb, Bi) are two novel titanium oxypnictides\cite{zhai2022structure}. Which are the first compounds with magnetic Eu$^{2+}$ incorporated. The structure of (EuF)$_2$Ti$_2$Pn$_2$O is almost identical to that of (SrF)$_2$Ti$_2$Pn$_2$O, being composed of alternating layers of Ti$_2$Pn$_2$O and fluorite-type EuF layers along the $c$-axis, as shown in Figure \ref{21}. The space group of (EuF)$_2$Ti$_2$Pn$_2$O are $I4/mmm$. The $a$-axis of (EuF)$_2$Ti$_2$Sb$_2$O is 4.106, the $c$-axis is 20.77, and the $a$-axis of (EuF)$_2$Ti$_2$Bi$_2$O is 4.117, the $c$-axis is 21.27, which is slightly smaller than (SrF)$_2$Ti$_2$Pn$_2$O. This is mainly due to the smaller atomic radius of Eu compared to Sr.

\begin{figure}
    \centering
    \includegraphics[scale=0.3]{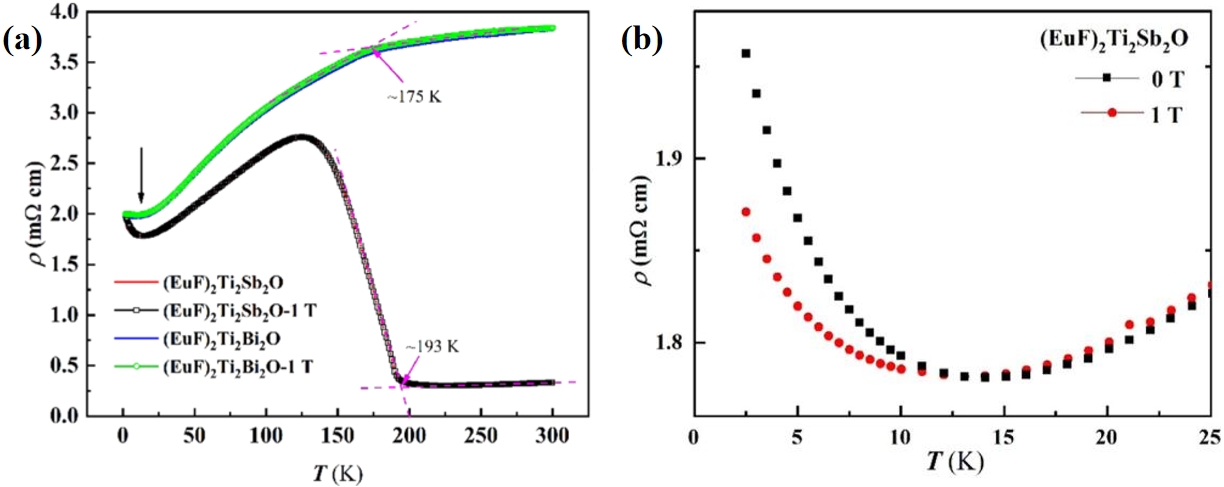}
    \caption{(a) Temperature dependence of resistivity measured under 0 T and 1 T for (EuF)$_2$Ti$_2$Pn$_2$O (Pn = Sb, Bi). (b) Temperature dependence of resistivity measured under 0 T and 1 T for (EuF)$_2$Ti$_2$Sb$_2$O in low-temperature regions\cite{zhai2022structure}.}
    \label{22}
\end{figure}

Figure \ref{22}a shows the resistivity of (EuF)$_2$Ti$_2$Pn$_2$O at 0 and 1 T. Above 200 K, we can see the resistivity is roughly constant and exhibits weak metallic behavior for (EuF)$_2$Ti$_2$Sb$_2$O. Near 193 K, it can be seen that a remarkably sudden growth of resistivity associated with a CDW/SDW instability. And in other titanium oxypnictides also observed similar substantial anomaly. Resistivity of (EuF)$_2$Ti$_2$Bi$_2$O has no obvious abnormality in the whole temperature range, only the change in the increasing rate of resistivity. It is worth noting that when the samples below 10 K, the resistivity will rise slightly at the marked by a black arrow in figure. The significant increase in resistivity suggests the formation of antiferromagnetic order among Eu local moments. The negative magnetoresistance behavior observed below 10 K in Figure \ref{22}b also supports the presence of magnetic order, which can be reasonably explained by the suppression of spin scattering in an external magnetic field. In addition, We can also find that the curves obtained by the samples under the magnetic fields of 0 T and 1 T are almost overlapped. This result may indicate that the contribution of magnetic scattering is negligible, unlike in iron-based compounds, where the antiferromagnetic-type SDW order is suppressed by an external magnetic field, leading to a distinct negative magnetoresistance\cite{dong2008competing}.

\begin{figure}
    \centering
    \includegraphics[scale=0.29]{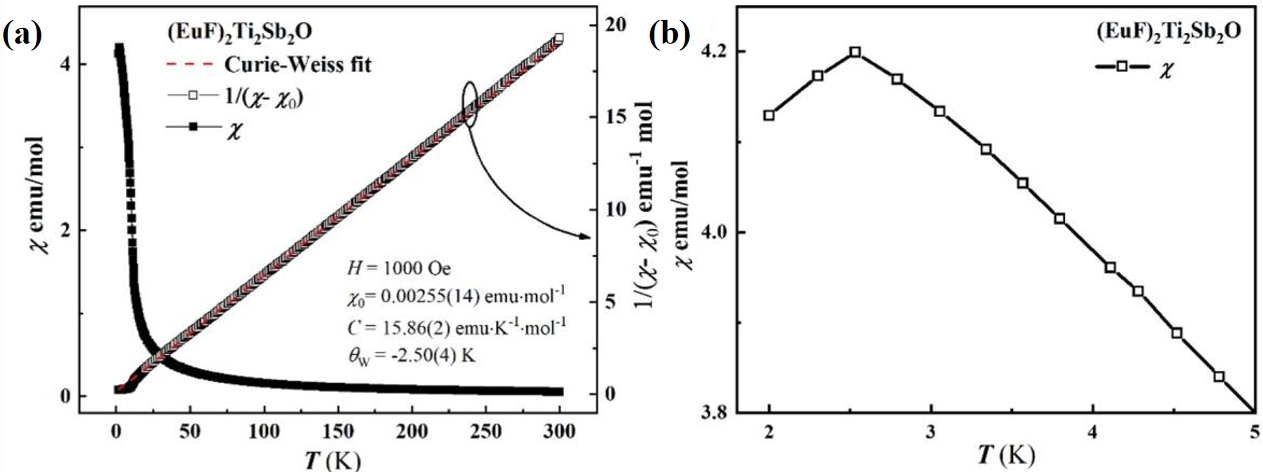}
    \caption{(a) Temperature dependence of magnetic susceptibility for (EuF)$_2$Ti$_2$Sb$_2$O, with the red line representing the Curie-Weiss fitting. (b) Temperature dependence of magnetic susceptibility (EuF)$_2$Ti$_2$Sb$_2$O in low-temperature regions\cite{zhai2022structure}.}
    \label{23}
\end{figure}

The magnetic susceptibility of (EuF)$_2$Ti$_2$Pn$_2$O show no significant abnormalities. This is in sharp contrast to other titanium pnictide oxides, whose magnetic susceptibility significantly decreases at the temperature of DW transition. The magnetic susceptibility of (EuF)$_2$Ti$_2$Sb$_2$O is shown in Figure \ref{23}a. The susceptibility data of the high-temperature part is very consistent with the extended Curie-Weiss law. However, in the low-temperature region, the fitted curve deviates from the Curie-Weiss law, and a peak appears at approximately 2.5 K as can be observed from Figure \ref{23}b. This abnormal phenomenon should correspond to the formation of antiferromagnetic ordering between Eu local moments. By calculating the magnetic moment of (EuF)$_2$Ti$_2$Pn$_2$O, it can be confirmed that all Eu ions in the compound are Eu$^{2+}$, but the magnetic moment in (EuF)$_2$Ti$_2$Bi$_2$O is slightly lower, which due to some Eu ions are freezing. It is worth noting that the susceptibility of (EuF)$_2$Ti$_2$Pn$_2$O is two orders of magnitude higher than other titanium pnictide oxides. This abnormally large susceptibility value may indicate hybridization between Eu-4f and conduction electrons, suggesting a strong electronic correlation in the system\cite{zhai2022structure}.

\begin{figure}
    \centering
    \includegraphics[scale=0.3]{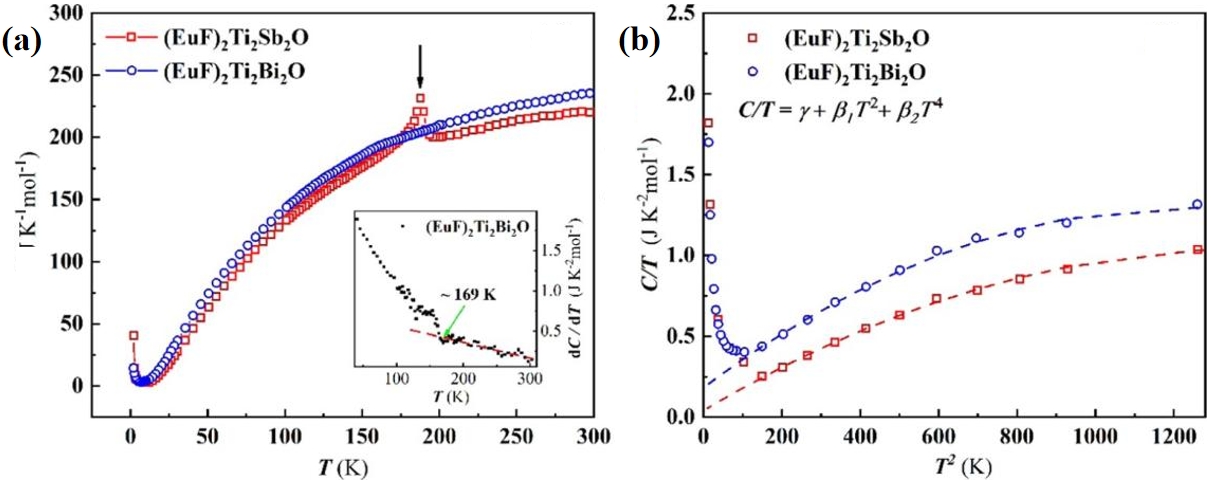}
    \caption{(a) Temperature dependence of magnetic susceptibility for (EuF)$_2$Ti$_2$Sb$_2$O, with the red line representing the Curie-Weiss fitting. (b) Temperature dependence of magnetic susceptibility (EuF)$_2$Ti$_2$Sb$_2$O in low-temperature regions\cite{zhai2022structure}.}
    \label{24}
\end{figure}

The specific heat capacity of (EuF)$_2$Ti$_2$Pn$_2$O, it can be seen in Figure \ref{24}a. The data tend to saturate at room temperature, consistent with the high-temperature limit for the lattice specific heat. It can be noted that at about 190 K, (EuF)$_2$Ti$_2$Sb$_2$O has an obvious peak. This is consistent with the temperature at which the anomalous behavior of resistivity occurs. For (EuF)$_2$Ti$_2$Bi$_2$O, only slight changes in slope were observed at high temperatures. From the illustration in Figure 23, it can be seen that the slope of (EuF)$_2$Ti$_2$Bi$_2$O changes significantly at around 169 K. At this temperature, the resistivity also undergoes a slope change (Figure \ref{22}a). These anomalies are highly likely due to DW behavior. In Figure \ref{24}b, the curve gives the Sommerfeld coefficient $\gamma$ = 41.98 mJ·K$^{-2}$·mol$^{-1}$, $\beta$ = 1.43 mJ·K$^{-4}$·mol$^{-1}$ for (EuF)$_2$Ti$_2$Sb$_2$O and $\gamma$ = 183.34 mJ·K$^{-2}$·mol$^{-1}$, $\beta$ = 1.81 mJ·K$^{-4}$·mol$^{-1}$ for (EuF)$_2$Ti$_2$Bi$_2$O. The Debye temperatures $\Theta_D$ of (EuF)$_2$Ti$_2$Sb$_2$O and (EuF)$_2$Ti$_2$Bi$_2$O are 230.2 K and 213.2 K respectively. It can be observed that $\gamma$ of (EuF)$_2$Ti$_2$Pn$_2$O is significantly higher than that of Na$_2$Ti$_2$Sb$_2$O and (SrF)$_2$Ti$_2$Bi$_2$O. 

For this phenomenon, a reasonable explanation is that the long-range magnetic ordering formation in the EuF layer has a significant impact on the Ti$_2$Pn$_2$O layer. This may mean that the electronic correlations between the neighboring EuF and Ti$_2$Pn$_2$O layers is significantly enhanced. And it can be observed that the DW behavior observed in (EuF)$_2$Ti$_2$Sb$_2$O is weaker than that in (SrF)$_2$Ti$_2$Sb$_2$O which may also an increase in carrier density of state at the Fermi surface of (EuF)$_2$Ti$_2$Sb$_2$O. However, the contributions of magnetic fluctuations cannot be completely ruled out.

Because these two compounds do not have magnetic ordering above 3 K, and the magnetic susceptibility follows well Curie-Weiss law above 15 K, the contribution from Eu magnetism is expected to be negligible at zero magnetic field above this temperature. 

(EuF)$_2$Ti$_2$Pn$_2$O are all non-superconducting. This may be due to the larger interlayer distance making the formation of Coulomb interlayer coupling more difficult, resulting in no superconducting transition occurring. This may be due to the larger interlayer distance making the formation of Coulomb interlayer coupling more difficult. However, the instability of CDW/SDW in (EuF)$_2$Ti$_2$Bi$_2$O is relatively weak, so regulating interlayer suppression of CDW/SDW instability may induce superconductivity. Although superconductivity is not observed in (EuF)$_2$Ti$_2$Pn$_2$O, europium-based layered compounds may provide new candidates for studying the relationships between multi-electron ordered states. The synthesis of such compounds may provide a new platform for studying the relationship between density waves and superconductivity

\section{Conclusions}
The recent unearthing of superconductivity within the novel stratified materials, notably the titanium-based pnictide oxides, has captured considerable attention. This interest stems from their amalgamation of traits reminiscent of both layered cuprates and iron-based superconductors. Pioneering investigations have documented CDW/SDW anomalies in titanium oxide (Ti$_2$O) compounds, which are instrumental in deciphering the intricate relationship between superconductivity and DW behavior.

Author offers a comprehensive dissection of the structural and physical attributes of titanium pnictide oxides, with a particular emphasis on their two-dimensional Ti$_2$O layers. These layers exhibit an anti-structural configuration relative to the CuO$_2$ planes and are capped by pnictogen ions. The compounds under scrutiny manifest CDW/SDW phases that coexist with superconductivity within certain family members, drawing parallels with iron pnictides but diverging from cuprates. Theoretical computations posit that these DW behaviors emanate from the Ti$_2$O layer, thereby laying the groundwork for exploring the nexus between superconductivity and DW behavior.

The synthesis and properties of an array of titanium pnictide oxides are deliberated in the review, encompassing Na$_2$Ti$_2$Pn$_2$O (Pn = As, Sb), (SrF)$_2$Ti$_2$Pn$_2$O (Pn = As, Sb, Bi), and (SmO)$_2$Ti$_2$Sb$_2$O. These compounds are noted for their CDW/SDW instabilities at elevated temperatures, yet they are devoid of superconductivity. Nevertheless, they offer invaluable insights into the structural and electronic interplays that may modulate superconducting properties.

The BaTi$_2$Pn$_2$O (Pn = As, Sb, Bi) compounds are accentuated for their superconducting properties, with BaTi$_2$Sb$_2$O and BaTi$_2$Bi$_2$O demonstrating superconductivity at 1.2 K and 4.6 K, respectively. Efforts to augment the critical temperature ($T_c$) of these compounds through doping have met with some success, enhancing their superconducting transition temperatures.

The Author also introduces (EuF)$_2$Ti$_2$Pn$_2$O (Pn = Sb, Bi) as emerging titanium oxypnictides incorporating magnetic Eu$^{2+}$. Despite the absence of superconductivity in these compounds, they may pave new pathways for probing the intricacies of multiple electronic ordering states.

In summary, there are many issues worth investigating in the future research of titanium pnictide oxide superconductors. It is essential to establish empirically and conceptually the nature of the DW phase to decipher its competition and synergy with the superconducting state. This type of new superconductor provides a new platform for explaining unconventional superconductivity. Titanium-based oxypnictides represent an intriguing class of layered materials, and future efforts are anticipated to undoubtedly advance our understanding of the interplay between superconductivity and DW behaviors, revealing their complex physical mechanisms.

\appendix
\section{Author contributions}
All authors have given approval to the final version of the manuscript.

\section{Conflicts of interest}
The authors declare no competing financial interests.




\bibliographystyle{elsarticle-num}

\end{document}